\documentclass[conference]{IEEEtran}
\usepackage{amsmath,amsfonts}
\usepackage{algorithmic}
\let\labelindent\relax
\usepackage{enumitem}
\usepackage{graphicx}
\usepackage{textcomp}
\usepackage{xcolor}
\def\BibTeX{{\rm B\kern-.05em{\sc i\kern-.025em b}\kern-.08em
    T\kern-.1667em\lower.7ex\hbox{E}\kern-.125emX}}
\setenumerate{itemsep=0pt}
\usepackage{hyperref}
\usepackage{cleveref}
\usepackage[normalem]{ulem}
\usepackage{xcolor}
\usepackage{pgfplots}
\usepackage{tikz}
\usepackage{subcaption}
\usepackage{algorithm}
\usepackage{makecell}

\pgfplotsset{compat=1.18}

\definecolor{bblue}{HTML}{4F81BD}
\definecolor{rred}{HTML}{C0504D}
\definecolor{ggreen}{HTML}{9BBB59}
\definecolor{ppurple}{HTML}{9F4C7C}
\definecolor{supergreen}{RGB}{0,150,0}

\newcommand{\mytilde}{\raisebox{0.5ex}{\texttildelow}}

\makeatletter
\newcommand\footnoteref[1]{\protected@xdef\@thefnmark{\ref{#1}}\@footnotemark}
\makeatother

\newcommand{\repository}{\url{https://github.com/FedericoMazzone/SmartCryptNN}}
\newcommand{\tool}{\textsc{SmartCryptNN}}

\DeclareMathOperator{\Z}{\mathbb{Z}}

\newcommand{\centralServer}{S}
\newcommand{\numParties}{N}
\newcommand{\party}{P}

\newcommand{\publicKey}{pk}
\DeclareMathOperator{\encrypt}{Encrypt}
\DeclareMathOperator{\decrypt}{Decrypt}

\DeclareMathOperator{\modelInit}{ModelInit}
\newcommand{\globalIt}{g}
\newcommand{\numGlobalIt}{E_g}
\newcommand{\localIt}{e}
\newcommand{\numLocalIt}{E_l}
\newcommand{\localBatchSize}{B}
\newcommand{\batchIndex}{b}
\newcommand{\model}{f}
\newcommand{\weight}{w}
\newcommand{\bias}{b}
\newcommand{\layer}{L}
\newcommand{\modelDepth}{L}
\newcommand{\modelArch}{a}
\newcommand{\modelActFunc}{\phi}
\newcommand{\learningCoeff}{\eta}
\newcommand{\decayCoeff}{\lambda}
\newcommand{\dataset}{D}

\newcommand{\privacyBudget}{\epsilon}
\newcommand{\clipGradRange}{\gamma}
\newcommand{\noise}{r}

\DeclareMathOperator{\laplacianDist}{Lap}
\newcommand{\encryptedLayers}{\mathcal{L}_S}
\newcommand{\exposedLayers}{\mathcal{L}_E}
\newcommand{\layerThreshold}{T}
\newcommand{\numExposedParams}{c}
\newcommand{\privacyThreshold}{\tau}

\makeatletter
\newenvironment{protocol}[1][htb]{%
    \renewcommand{\ALG@name}{Protocol}
    \begin{algorithm}[#1]%
    }{\end{algorithm}
}
\makeatother

\begin{document}
\title{Investigating Privacy Attacks in the Gray-Box Setting to Enhance Collaborative Learning Schemes}

\author{
\IEEEauthorblockN{Federico Mazzone}
\IEEEauthorblockA{University of Twente\\
f.mazzone@utwente.nl}
\\
\IEEEauthorblockN{Maarten Everts}
\IEEEauthorblockA{University of Twente\\Linksight\\
maarten.everts@utwente.nl}
\and
\IEEEauthorblockN{Ahmad Al Badawi}
\IEEEauthorblockA{Duality Technologies\\
aalbadawi@dualitytech.com}
\\
\IEEEauthorblockN{Florian Hahn}
\IEEEauthorblockA{University of Twente\\
f.w.hahn@utwente.nl}
\and
\IEEEauthorblockN{Yuriy Polyakov}
\IEEEauthorblockA{Duality Technologies\\
ypolyakov@dualitytech.com}
\\
\IEEEauthorblockN{Andreas Peter}
\IEEEauthorblockA{University of Oldenburg\\
andreas.peter@uni-oldenburg.de}
}

\maketitle

\begin{abstract}
The notion that collaborative machine learning can ensure privacy by just withholding the raw data is widely acknowledged to be flawed.
Over the past seven years, the literature has revealed several privacy attacks that enable adversaries to extract information about a model's training dataset by exploiting access to model parameters during or after training.

In this work, we study privacy attacks in the gray-box setting, where the attacker has only limited access --- in terms of view and actions --- to the model.
The findings of our investigation provide new insights for the development of privacy-preserving collaborative learning solutions.
We deploy \tool, a framework that tailors homomorphic encryption to protect the portions of the model posing higher privacy risks.

Our solution offers a trade-off between privacy and efficiency, which varies based on the extent and selection of the model components we choose to protect.
We explore it on dense neural networks, where through extensive evaluation of diverse datasets and architectures, we uncover instances where a favorable sweet spot in the trade-off can be achieved by safeguarding only a single layer of the network.
In one of such instances, our approach trains \mytilde4 times faster compared to fully encrypted solutions, while reducing membership leakage by 17.8 times compared to plaintext solutions.
\end{abstract}

\section{Introduction}
\label{sec:intro}

The demand for more complex and accurate machine learning models in fields like image recognition and natural language processing has highlighted the need for extensive training data~\cite{zhu2012we}.
As a result, collaboration among data-owning entities has become crucial to leverage larger and more diverse datasets and enhance model performance.
However, this collaborative approach raises privacy concerns, as sharing raw data can expose sensitive information, potentially violating privacy regulations and raising confidentiality concerns (for instance, this may be the case for hospitals, municipalities, insurance companies, etc.).
To address this issue, collaborative learning solutions, like Federated Learning (FL)~\cite{mcmahan2017communication} and Split Learning (SL)~\cite{poirot2019split}, have been proposed.
Those methods aim to protect privacy by enabling the training of a joint model without directly outsourcing the raw data.

The recent development of privacy attacks has made evident that solely withholding the training data is an insufficient strategy to ensure privacy protection in these scenarios~\cite{nasr2019comprehensive, melis2019exploiting, fredrikson2015model,hitaj2017deep,zhu2019deep}.
These attacks aim to retrieve information about the dataset a given model has been trained on by only having access to the model parameters (\textit{white-box setting}), or to the associated prediction functionality (\textit{black-box setting}).
To mitigate such attacks, solutions based on Differential Privacy (DP)~\cite{shokri2015privacy,abadi2016deep,mcmahan2017learning}, Fully Homomorphic Encryption (FHE)~\cite{sav2020poseidon}, and MultiParty Computation (MPC)~\cite{wagh2020falcon,mohassel2018aby3,wagh2019securenn} have been introduced.
However, these solutions come with trade-offs: DP introduces noise to protect privacy but may lead to accuracy loss; FHE provides strong privacy guarantees but has a high computation cost for deep arithmetic circuits that require bootstrapping, making it an unfeasible approach in many practical training contexts; MPC may require a large number of interactions between multiple parties and high communication bandwidth.

In this paper, we go beyond the conventional white- and black-box settings commonly explored in existing Privacy-Preserving Machine Learning (PPML) literature~\cite{nasr2019comprehensive},
and we explore a more flexible \textit{gray-box setting} where the model is only \textit{partially exposed} to the adversary.
We investigate different classes of privacy attacks
and analyze how the efficacy of the attacks is influenced by the limitations imposed on the adversary's view and actions.
For neural networks, which represent our main research focus,
this gray-box setting can manifest as an attacker having access only to specific layers of the model, rather than the entire architecture.
This particular setting occurs naturally in scenarios like SL, where the server only has access to the second split of the model.
It is also relevant for non-collaborative scenarios, like the deployment of generative models: think for instance about the training of an AutoEncoder or a GAN, where only the decoder or the generator, respectively, is deployed to the public.
From a privacy evaluation perspective, the gray-box setting can be seen as a generalization of such scenarios.

While prior work in the literature indirectly initiated the analysis of the gray-box setting for specific attacks or scenarios~\cite{nasr2019comprehensive,pasquini2021unleashing}, our aim is to provide a more comprehensive and methodical investigation.
To this end, we consider multiple classes of privacy attacks, including membership inference, property inference, and model inversion~\cite{rigaki2020survey}.
By adapting state-of-the-art attacks to suit the gray-box setting, we assess their efficacy via both theoretical considerations and empirical evaluation.
Our findings reveal that certain classes of attacks exhibit greater performance when specific layers of the model are accessible.
For instance, membership inference attacks show a higher sensitivity towards the latter layers of the model, while model inversion techniques heavily rely on the availability of the initial layers.

\begin{figure}
    \captionsetup[subfigure]{}
    \centering
    \begin{subfigure}[t]{.5\columnwidth}
        \centering
        \includegraphics[width=0.78\columnwidth]{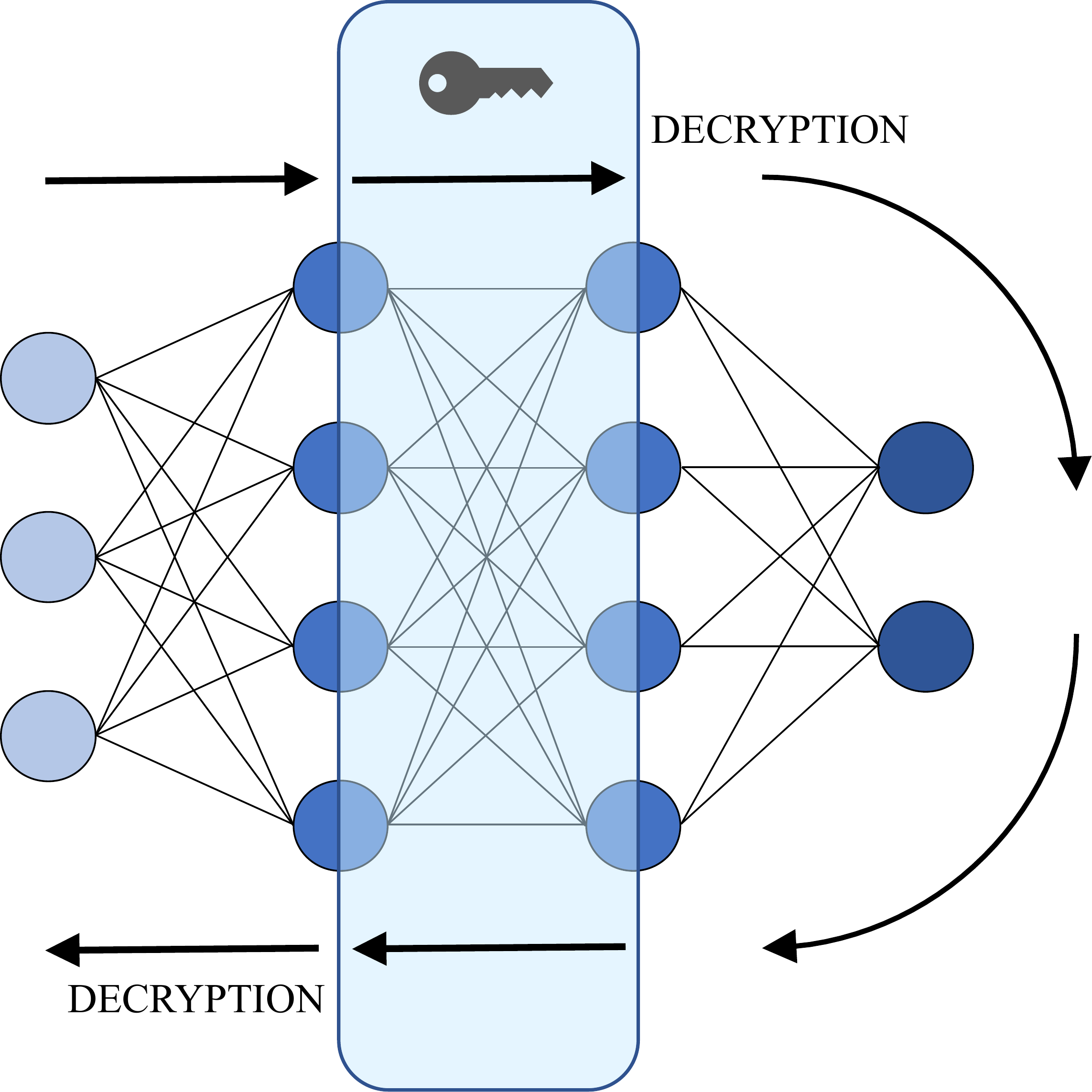}
        \caption{Hidden layer encryption.}
        \label{fig:partially_encrypted_nn:central_layer}
    \end{subfigure}%
    \begin{subfigure}[t]{.5\columnwidth}
        \centering
        \includegraphics[width=0.95\columnwidth]{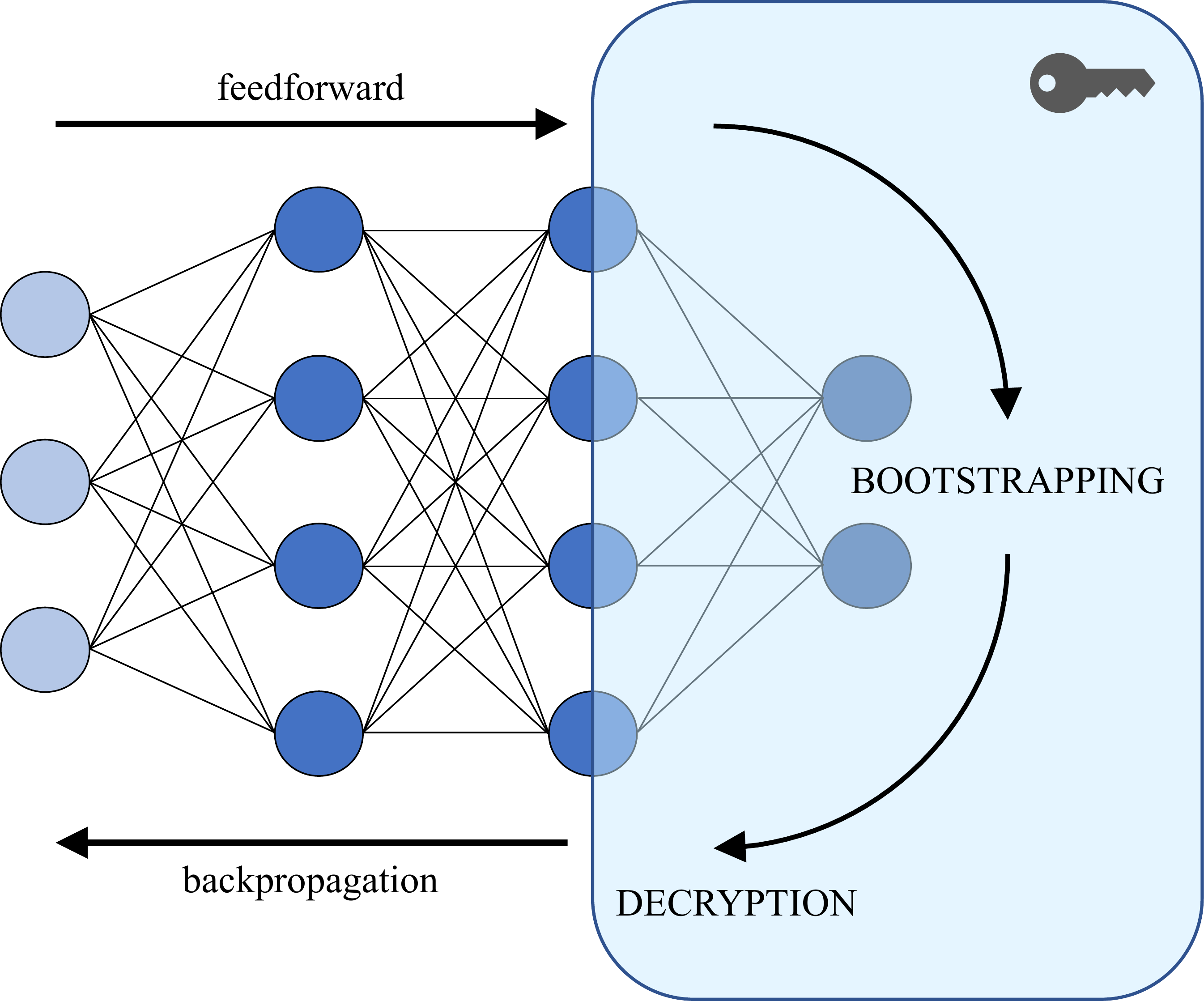}
        \caption{Output layer encryption.}
        \label{fig:partially_encrypted_nn:last_layer}
    \end{subfigure}
    \caption{Diagram representation of our approach.}
    \label{fig:partially_encrypted_nn}
\end{figure}

Building upon these insights, we show how to leverage them to enhance collaborative learning solutions in terms of privacy, efficiency, and accuracy.
We adopt a targeted defense strategy by applying homomorphic encryption to safeguard only the most vulnerable portions of the model (see \Cref{fig:partially_encrypted_nn}).
This way, the resulting gradient updates align with the encryption status of the layers, ensuring that encrypted layers remain encrypted and the exposed layers remain in plaintext throughout the entire training process.
The resulting approach offers higher privacy guarantees compared to plain FL, while concurrently achieving less computational and communication overhead when compared to fully encrypted solutions~\cite{sav2020poseidon}.
The freedom to choose the number of layers to protect introduces a novel trade-off between privacy and efficiency.
In practical scenarios, this enables the final user to flexibly fine-tune the solution to meet specific efficiency requirements while sacrificing little privacy.
Our investigation reveals that in certain instances encrypting only a few layers of the model is indeed enough to strongly inhibit most privacy attacks.
For instance, encrypting just one layer was enough to reduce a membership inference leakage by 17.8 times.
Notably, our approach becomes particularly efficient for deep models.

Similar to other work~\cite{sav2020poseidon}, we only consider the semi-honest setting.
In \Cref{sec:pems:security}, we see that active variants of privacy attacks are not feasible in this case, hence we exclude them from our in-depth analysis.
Techniques developed against Split Learning, such as the ones described in~\cite{pasquini2021unleashing}, usually focus on model inversion in the malicious setting, and as such are not part of our investigation.
A detailed description of our threat model can be found in \Cref{sec:threat_model}.

We carry out our investigation on diverse datasets for supervised classification tasks, spanning both tabular and image features.
On those datasets, we assess both the privacy and efficiency of our partially encrypted FL strategy, observing how such a trade-off evolves based on the specific portion of the model that is protected.
To empirically measure the computation and communication costs of our approach, we implement it in a framework named \tool, a prototype built upon the multiparty CKKS-RNS variant provided by the OpenFHE library~\cite{albadawi2022openfhe}.
Additionally, to ensure consistency with the work by Sav et al.~\cite{sav2020poseidon}, we simulate the protocol communication using MiniNet, allowing us to set different network delay and bandwidth constraints.
To the best of our knowledge, we are the first to release an open-source implementation of FL under homomorphic encryption.

The main contributions of this work are listed below:
\begin{itemize}
    \item We adapt different classes of privacy attacks to the gray-box setting and we assess how their efficacy scales with the portion of the model available to the adversary.
    \item We design a privacy-preserving collaborative training solution for deep learning that selectively encrypts only the layers of the model with higher privacy risk.
    \item We implement our solution and assess it on a variety of well-known datasets, making our framework available open-source at \repository.
\end{itemize}

\section{Background}
\label{sec:background}

In this section, we provide a brief description of some basic ML and PPML-related concepts and introduce the models and datasets used in our investigation.

\subsection{Federated Learning}
Federated Learning (FL) is a collaborative learning approach where multiple data-owning clients, $\party_1, \dots, \party_\numParties$, jointly train a common model while keeping their data decentralized.
Originally, Shokri and Shmatikov~\cite{shokri2015privacy} introduced an FL scheme for neural networks where clients independently train local models on their datasets, while sharing portions of their model parameters with a global model hosted by a supporting server. The training process is asynchronous, with each client repeatedly downloading portions of the global model parameters, updating its local model, and then uploading portions of the gradients to the global model.

However, a more widely adopted approach for FL is Federated Averaging (FedAvg), as proposed by McMahan et al.~\cite{mcmahan2017communication}. In FedAvg, the training process occurs in synchronous rounds. In each round, a subset of clients is selected to participate. These clients perform a local training step on their own data, updating their model parameters. These locally trained models are then sent to the central server, where they are averaged together to obtain a new global model. The aggregated global model is then distributed back to all participating clients.

\subsection{Multilayer Perceptrons}
In this work, we focus on Multilayer Perceptrons (MLPs), for which we provide some notation as follows.
For an MLP model $\model$ with $\modelDepth$ layers, we denote by $\weight_i$, $\bias_i$, $\modelActFunc_i$ the weights, biases, and activation function of layer $i$, respectively.
We denote by $l_i$ the output of layer $i$, that is $l_i = \modelActFunc_i(l_{i-1} \weight_i + \bias_i)$, where $l_0 := x$ is the input of the model.
And we denote the intermediate linear application output as $u_i = l_{i-1} \weight_i + \bias_i$.
For supervised classification tasks, the loss function is denoted by $L(\model(x), y)$, where $y$ is the ground-truth label associated to $x$, and can be minimized through Gradient Descent (GD) techniques.
Below is a schematic representation of the computations performed during one step of the training process.
More details can be found in \Cref{app:mlp}.

\begin{figure}[!htb]
    \centering
    \resizebox{!}{111pt}{%
    \begin{tikzpicture}[]
        \node (input) at (0.1, 0) {$l_0 = x$};
        \node[anchor=south] (output1) at (1, 1.2) {$\begin{aligned} u_1 &= l_0 \weight_1 + \bias_1 \\ l_1 &= \modelActFunc_1(u_1) \end{aligned}$};
        \node[anchor=south] (outputDots) at (3.5, 1.6) {$\cdots$};
        \node[anchor=south] (outputL) at (6, 1.2) {$\begin{aligned} u_\modelDepth &= l_{\modelDepth-1} \weight_\modelDepth + \bias_\modelDepth \\ l_\modelDepth &= \modelActFunc_\modelDepth(u_\modelDepth) \end{aligned}$};
        \node[align=center] (loss) at (6.9, 0) {$L(l_\modelDepth, y)$ \\ $e_\modelDepth = \partial L / \partial l_\modelDepth$};
        \node[align=center, anchor=north] (errL) at (6, -1.2) {$\begin{aligned} \nabla \bias_\modelDepth &= e_\modelDepth \modelActFunc'_\modelDepth(u_\modelDepth) \\ \nabla \weight_\modelDepth &= e_\modelDepth \modelActFunc'_\modelDepth(u_\modelDepth) l_{\modelDepth-1}^T \\ e_{\modelDepth-1} &= e_\modelDepth \modelActFunc'_\modelDepth(u_\modelDepth) \weight_\modelDepth^T \end{aligned}$};
        \node[anchor=north] (errDots) at (3.5, -1.8) {$\cdots$};
        \node[align=center, anchor=north] (err1) at (1, -1.2) {$\begin{aligned} \nabla \bias_1 &= e_1 \modelActFunc'_1(u_1) \\ \nabla \weight_1 &= e_1 \modelActFunc'_1(u_1) l_0^T \\ \big( e_0 &= e_1 \modelActFunc'_1(u_1) \weight_1^T \big) \end{aligned}$};
        \draw[->, shorten <=4pt, shorten >=4pt] (input) -- (output1);
        \draw[->] (2.6, 1.8) -- (3.1, 1.8);
        \draw[->] (3.8, 1.8) -- (4.2, 1.8);
        \draw[->, shorten <=4pt, shorten >=4pt] (outputL) -- (loss);
        \draw[->, shorten <=2pt, shorten >=4pt] (loss) -- (errL);
        \draw[->] (4.2, -2.0) -- (3.8, -2.0);
        \draw[->] (3.1, -2.0) -- (2.6, -2.0);
    \end{tikzpicture}
    }
    \label{fig:one_pass}
\end{figure}

\subsection{Multiparty Fully Homomorphic Encryption}
Fully Homomorphic Encryption (FHE) enables the evaluation of unlimited-depth arithmetic circuits on encrypted data, utilizing a technique called bootstrapping for refreshing ciphertexts after homomorphic operations.
In Multiparty Homomorphic Encryption (MHE), the secret key is shared among multiple parties, who can use their shares to distributively generate collective public/evaluation keys and perform distributed decryption and bootstrapping protocols.
For our work, we use the threshold FHE version of the Cheon-Kim-Kim-Song (CKKS) scheme~\cite{cheon2017homomorphic}. The threshold FHE construction follows the design of Asharov et al.~\cite{asharov2012multiparty}, which was initially instantiated for the Brakerski-Gentry-Vaikuntanathan (BGV) scheme~\cite{BGV}. Mouchet et al.~\cite{mouchet2021multiparty} applied this blueprint to the Brakerski / Fan-Vercauteren (BFV)~\cite{Bra12,FV12} and CKKS schemes, also introducing a distributed bootstrapping protocol based on threshold FHE. The distributed CKKS bootstrapping protocol was further developed in Sav et al.~\cite{sav2020poseidon}.
Additional information about threshold CKKS can be found in \Cref{app:threshold_CKKS}.

To instantiate threshold CKKS, we use the OpenFHE library~\cite{albadawi2022openfhe}, which implements optimized CKKS variants proposed in~\cite{kim2022approximate} and threshold FHE extensions for CKKS.
We use a $N$-out-of-$N$ threshold scheme with additive sharing of the secret key, where all parties need to be present to perform decryption and bootstrapping. But the scheme can be easily modified to allow for arbitrary thresholds, i.e., $t$-out-of-$N$ threshold FHE using Shamir secret sharing.

The CKKS scheme is well-suited for floating-point-like arithmetic and performs computations on vectors of real/complex numbers, allowing for Single Instruction, Multiple Data (SIMD) operations.
In particular, the scheme works with residual polynomial rings of the form $R_q = \Z_q[x] / (x^n + 1)$, where $n$ is a power of two.
Note that due to the CKKS encoding, a plaintext can embed a vector of up to $n / 2$ slots.

We use the alternate packing approach designed by Sav et al.\cite{sav2020poseidon} to encode matrices, and Chebyshev interpolation to approximate non-polynomial functions (more details in \Cref{app:he_operations}).

We work with a Residue Number System (RNS) instantiation of CKKS~\cite{kim2022approximate}, which achieves the highest efficiency for CKKS among known variants of the scheme. Here, we briefly describe it at a high level. Given unique primes $q_0, q_1, \dots, q_\modelDepth$, an RNS chain of moduli is built as $Q_i = \prod_{j=0}^i{q_j}$ for $i \in \{0, \dots, L\}$.
A freshly encrypted ciphertext is a pair $(c_0, c_1) \in R_{Q_L}$. Then, after each multiplication, the ciphertext is rescaled to scale down the message and truncate the least significant bits, dropping the highest RNS limb, e.g., going from $R_{Q_L}$ to $R_{Q_{L-1}}$ after first rescaling. The maximum number of multiplications is given by $L-1$. However, not all of these levels can be used for the main computation as the bootstrapping procedure consumes levels, too (even in the case of distributed bootstrapping).

The main drawbacks of MHE that often make it impractical in real-world contexts are the heavy computation costs and significant communication overhead, especially due to the need for distributed bootstrapping, which can become a bottleneck in practical implementations. This is particularly relevant in ML scenarios when evaluating deep multiplicative circuits like neural networks.
Additionally, the inherent noise in the encryption scheme and the approximation of non-linear functions can lead to a decrease in the model's accuracy.
To address these challenges and enhance both efficiency and accuracy, our work focuses on reducing the number of layers that need to be encrypted.

\subsection{Datasets and Models}
\label{sec:background:data_and_models}
We selected well-known public datasets widely used in the PPML literature: with Texas-100, Purchase-100, Locations our datasets include tabular data, and with AT\&T, MNIST, EMNIST Letters, LFW our datasets include images.
We refer to \Cref{app:dataset} for further details. 
This mix ensures a comprehensive evaluation of the privacy attacks and our prototype.
We highlight that this work's main goal is not to achieve the highest possible accuracy on the given datasets, but to investigate the efficacy of privacy attacks across different layers of our target model.
To accomplish this, we deliberately subsampled some of the datasets, using a reduced dataset for training the models. By doing so, we aim to create a vulnerable model that is more susceptible to privacy attacks. As discussed in \Cref{sec:attacks} there are various ways to make a model vulnerable.
Constraining the training set to a subset is compatible with real-world scenarios, where training parties might struggle with limited data availability.
This approach also helps in expediting the experimental assessment of our encrypted training solution, as the experiments can be run within a reasonable timeframe given the reduced training data size.

As for the models, we primarily focus on MLP architectures, for compatibility with our FHE prototype, using the plain Stochastic Gradient Descent (SGD) optimizer and minimizing the Mean Squared Error (MSE) loss.
Specifically, we train an MLP with two hidden layers of size 30 and 20 on the MNIST datasets, and 256, 128, and 64 on the Location dataset.
For Purchase100 and Texas100, we adopt the MLP architecture proposed in~\cite{nasr2019comprehensive}, which consists of hidden layers with sizes 1024, 512, 256, and 128.
Additionally, we train an MLP with two hidden layers of size 64 on the EMNIST Letters dataset, and a CNN on the LFW dataset.
For the CNN, we adopt the architecture proposed in~\cite{melis2019exploiting}, which consists of three convolutional
layers with 32, 64, and 128 filters, each with a 3x3 kernel and a max pooling layer, followed by two fully connected layers of size 256 and 2.
For further details about the training settings, we refer to the repository.\footnote{\repository}.

\section{Threat Model}
\label{sec:threat_model}
In this section, we briefly discuss our considered threat models.  
We distinguish between the information available to the adversary and the actions performed by the adversary. 

\subsection{Adversary's View}
\label{sec:threat_model:adv_view}
PPML literature classifies privacy attacks depending on the adversary's view of the model~\cite{nasr2019comprehensive}.
In black-box attacks, the adversary can only access the model's output for arbitrarily chosen inputs but lacks information about model parameters.
In white-box attacks, the adversary has full access to the model's architecture, parameters, and hyperparameters used during training.
This enables them to compute any function of the model parameters and any chosen input, including intermediate computations of the feedforward pass, i.e.~output of intermediate layers.
For labeled input, the adversary can hence compute the corresponding loss and the gradients for each layer.

We study attacks in the \textit{gray-box setting}, which represents a more flexible threat model generalizing white-box scenarios for privacy-attacks, by considering intermediate adversary capabilities.
In this gray-box setting, the adversary has only access to a subset of the model parameters and can compute any function limited to those parameters.
Specifically, for the studied feed-forward neural networks, we consider a layer-wise granularity for this partial knowledge.
This entails partitioning the model's layers into two subsets: the \emph{exposed layers} $\exposedLayers$ and the \emph{secret layers} $\encryptedLayers$.
For the exposed layers, the adversary has the same access as in the white-box setting.
Thus, the adversary can compute the loss value for a given labeled example only if the last layer is exposed.
Conversely, the adversary has no access to any parameters of secret layers.

\subsection{Adversary's Actions}
\label{sec:threat_model:adv_actions}
In the scenarios where the adversary joins the training phase, PPML literature distinguishes between passive and active behaviour~\cite{nasr2019comprehensive,melis2019exploiting} for privacy attacks.
However, this taxonomy is inconsistent across prior works, and can cause confusion with terminology used in cryptography. 
Since in this paper we address both the ML and cryptographic aspects, we clarify our threat model as follows.

From a machine learning perspective, we closely follow the definition given by Nasr et al.~\cite{nasr2019comprehensive} and distinguish between \textit{ML-passive} and \textit{ML-active} adversaries.
An ML-passive adversary only observes passively the legitimate model updates and attempts to infer information by performing inference on the model, without changing anything in the local or global collaborative training procedure.
In contrast, an ML-active adversary, influences the target model during training in order to coerce the data owners into unintentionally releasing more information through the model. 
The active adversary's actions may include choosing specific artificially crafted inputs
for the training procedure, or performing gradient ascents on specific inputs.
We specifically use this definition in \Cref{sec:attacks}.

Note that this distinction differs from the cryptographic definition of passive (or semi-honest) and active (or malicious) adversaries.
A \textit{crypto-passive} adversary follows the protocol, while a \textit{crypto-active} adversary can arbitrarily deviate from the protocol.
This cryptographic definition is more powerful than the machine learning definition of an active adversary, as the latter only allows for local manipulation of the model.
In the cryptographic sense, a passive adversary is potentially allowed to feed arbitrary input to the model training procedure, but must adhere to the specified protocol.
We specifically use this definition in \Cref{sec:pems}.

\section{Privacy Attacks in the Gray-Box Setting}
\label{sec:attacks}

In this section, we assess existing privacy attacks in the gray-box setting, aiming to get insights into what portions of the model are more vulnerable to different kinds of attacks.
In general, a privacy attack is a technique designed to extract information about the data that a particular model has been trained on.
Privacy attacks are typically categorized based on the specific type of information they aim to extract.
Following the taxonomy provided by Rigaki et al. in their survey~\cite{rigaki2020survey}, we classify privacy attacks into three categories: membership inference, model inversion, and property inference.
To provide a comprehensive evaluation of privacy attacks, we considered all the aforementioned classes in our investigation.
For each of these classes, we discuss how the attacks can operate in the gray-box setting described in~\Cref{sec:threat_model}, and evaluate their effectiveness based on the degree of exposure of the model to the adversary.
We select state-of-the-art attacks from the existing literature to experimentally support our assessment.
These attacks often come with an ML-active variant, whose efficacy in the gray-box setting is highly dependent on how the secret layers are concealed and the limitations imposed on the adversary's actions.
We discuss the availability of their ML-active counterparts directly in \Cref{sec:pems:security}.

\subsection{Membership Inference}
\label{sec:attacks:memb_inf}

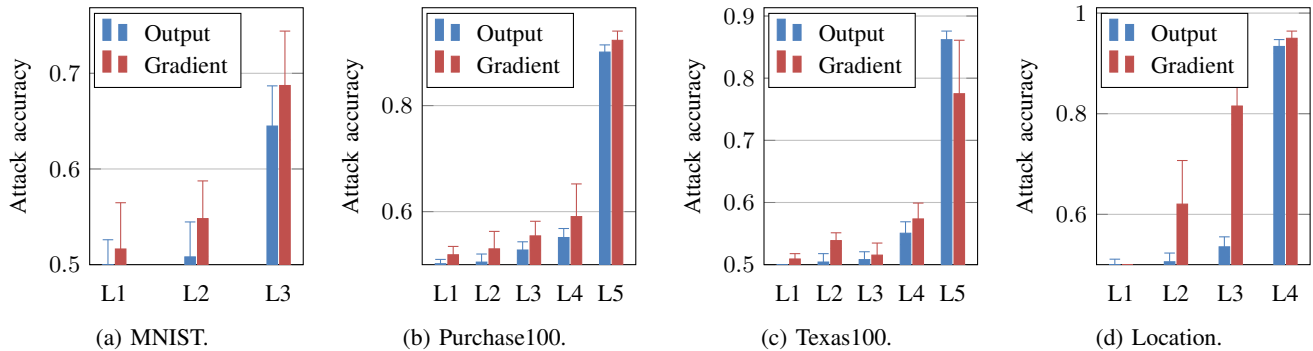
\begin{figure*}[!htb]
    \captionsetup[subfigure]{}
    \centering
    \begin{subfigure}{.24\textwidth}
        \centering
        \begin{tikzpicture}
            \tikzstyle{every node}=[font=\small]
            \begin{axis}[
                width = 1.0*\linewidth,
                height = 5cm,
                major x tick style = transparent,
                ybar=2*\pgflinewidth,
                bar width=4pt,
                ymajorgrids = true,
                ylabel = {Attack accuracy},
                symbolic x coords={L1, L2, L3},
                xtick = data,
                scaled y ticks = false,
                enlarge x limits=0.15,
                ymin=0.5,
                legend cell align=left,
                legend style={
                        at={(0.02,0.98)},
                        anchor=north west,
                        column sep=1ex
                }
            ]
                \addplot[style={bblue,fill=bblue,mark=none}, error bars/.cd, y dir=both, y explicit relative]
                    coordinates {
                        (L1, 0.49299997091293335) +- (0,0.06700003147125244)
                        (L2, 0.5081250071525574) +- (0,0.07187497615814209)
                        (L3, 0.6448749899864197) +- (0,0.0651249885559082)
                    };
                \addplot[style={rred,fill=rred,mark=none}, error bars/.cd, y dir=both, y explicit relative]
                     coordinates {
                        (L1, 0.5163750201463699) +- (0,0.09362499415874481)
                        (L2, 0.5481250137090683) +- (0,0.07187499105930328)
                        (L3, 0.6872499883174896) +- (0,0.08274999260902405)
                    };
                \legend{Output, Gradient}
            \end{axis}
        \end{tikzpicture}
        \caption{
            MNIST.
        }
        \label{fig:memb_inf_eval:MNIST}
    \end{subfigure}%
    \begin{subfigure}{.24\textwidth}
        \centering
        \begin{tikzpicture}
            \tikzstyle{every node}=[font=\small]
            \begin{axis}[
                width = 1.0*\linewidth,
                height = 5cm,
                major x tick style = transparent,
                ybar=2*\pgflinewidth,
                bar width=4pt,
                ymajorgrids = true,
                ylabel = {Attack accuracy},
                symbolic x coords={L1, L2, L3, L4, L5},
                xtick = data,
                scaled y ticks = false,
                enlarge x limits=0.15,
                ymin=0.5,
                legend cell align=left,
                legend style={
                        at={(0.02,0.98)},
                        anchor=north west,
                        column sep=1ex
                }
            ]
                \addplot[style={bblue,fill=bblue,mark=none}, error bars/.cd, y dir=both, y explicit relative]
                    coordinates {
                        (L1, 0.5017375349998474) +- (0,0.016262471675872803)
                        (L2, 0.5048249363899231) +- (0,0.030175089836120605)
                        (L3, 0.5276249796152115) +- (0,0.029375001788139343)
                        (L4, 0.5512000322341919) +- (0,0.0307999849319458)
                        (L5, 0.9009625762701035) +- (0,0.015037432312965393)
                    };
                \addplot[style={rred,fill=rred,mark=none}, error bars/.cd, y dir=both, y explicit relative]
                     coordinates {
                        (L1, 0.5186499953269958) +- (0,0.030350029468536377)
                        (L2, 0.5298000574111938) +- (0,0.062199950218200684)
                        (L3, 0.5542749911546707) +- (0,0.04972498118877411)
                        (L4, 0.5906375050544739) +- (0,0.10436248779296875)
                        (L5, 0.9231375306844711) +- (0,0.018862441182136536)
                    };
                \legend{Output, Gradient}
            \end{axis}
        \end{tikzpicture}
        \caption{
            Purchase100.
        }
        \label{fig:memb_inf_eval:purchase100}
    \end{subfigure}
    \begin{subfigure}{.24\textwidth}
        \centering
        \begin{tikzpicture}
            \tikzstyle{every node}=[font=\small]
            \begin{axis}[
                width = 1.0*\linewidth,
                height = 5cm,
                major x tick style = transparent,
                ybar=2*\pgflinewidth,
                bar width=4pt,
                ymajorgrids = true,
                ylabel = {Attack accuracy},
                symbolic x coords={L1, L2, L3, L4, L5},
                xtick = data,
                scaled y ticks = false,
                enlarge x limits=0.15,
                ymin=0.5,
                legend cell align=left,
                legend style={
                        at={(0.02,0.98)},
                        anchor=north west,
                        column sep=1ex
                }
            ]
                \addplot[style={bblue,fill=bblue,mark=none}, error bars/.cd, y dir=both, y explicit relative]
                    coordinates {
                        (L1, 0.4927250146865845) +- (0,0.007274985313415527)
                        (L2, 0.5042250156402588) +- (0,0.026775002479553223)
                        (L3, 0.5081125497817993) +- (0,0.024887442588806152)
                        (L4, 0.5505749583244324) +- (0,0.03342503309249878)
                        (L5, 0.8619249910116196) +- (0,0.016075029969215393)
                    };
                \addplot[style={rred,fill=rred,mark=none}, error bars/.cd, y dir=both, y explicit relative]
                     coordinates {
                        (L1, 0.5091375112533569) +- (0,0.01686251163482666)
                        (L2, 0.5387125164270401) +- (0,0.023287460207939148)
                        (L3, 0.5153625011444092) +- (0,0.03763747215270996)
                        (L4, 0.5735749751329422) +- (0,0.04442499577999115)
                        (L5, 0.7749999761581421) +- (0,0.11100000143051147)
                    };
                \legend{Output, Gradient}
            \end{axis}
        \end{tikzpicture}
        \caption{
            Texas100.
        }
        \label{fig:memb_inf_eval:texas100}
    \end{subfigure}%
    \begin{subfigure}{.24\textwidth}
        \centering
        \begin{tikzpicture}
            \tikzstyle{every node}=[font=\small]
            \begin{axis}[
                width = 1.0*\linewidth,
                height = 5cm,
                major x tick style = transparent,
                ybar=2*\pgflinewidth,
                bar width=4pt,
                ymajorgrids = true,
                ylabel = {Attack accuracy},
                symbolic x coords={L1, L2, L3, L4},
                xtick = data,
                scaled y ticks = false,
                enlarge x limits=0.15,
                ymin=0.5,
                legend cell align=left,
                legend style={
                        at={(0.02,0.98)},
                        anchor=north west,
                        column sep=1ex
                }
            ]
                \addplot[style={bblue,fill=bblue,mark=none}, error bars/.cd, y dir=both, y explicit relative]
                    coordinates {
                        (L1, 0.5001041665673256) +- (0,0.021562479436397552)
                        (L2, 0.5057396292686462) +- (0,0.03426039218902588)
                        (L3, 0.5354479551315308) +- (0,0.037052035331726074)
                        (L4, 0.9336979240179062) +- (0,0.014635398983955383)
                    };
                \addplot[style={rred,fill=rred,mark=none}, error bars/.cd, y dir=both, y explicit relative]
                     coordinates {
                        (L1, 0.5) +- (0,0.0)
                        (L2, 0.620343804359436) +- (0,0.1396561861038208)
                        (L3, 0.8152916580438614) +- (0,0.04554168879985809)
                        (L4, 0.9499271064996719) +- (0,0.015072867274284363)
                    };
                \legend{Output, Gradient}
            \end{axis}
        \end{tikzpicture}
        \caption{
            Location.
        }
        \label{fig:memb_inf_eval:location}
    \end{subfigure}
    \caption{Layer-wise accuracy of the white-box membership inference attack by Nasr et al.~\cite{nasr2019comprehensive} against different datasets and models, exploiting both the layer's output and gradient.
    }
    \label{fig:memb_inf_eval}
\end{figure*}

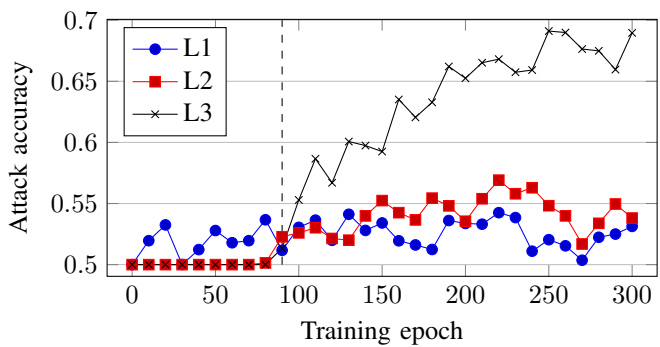
\begin{figure}
    \centering
    \begin{tikzpicture}
        \begin{axis}[
            height=5cm,
            width=\columnwidth,
            xlabel={Training epoch},
            ylabel={Attack accuracy},
            enlarge x limits=0.05,
            enlarge y limits=0.05,
            legend pos=north west,
            ymajorgrids=true,
        ]
        
        \addplot
            coordinates {
            (0, 0.5000)
            (10, 0.5196)
            (20, 0.5325)
            (30, 0.5001)
            (40, 0.5123)
            (50, 0.5279)
            (60, 0.5179)
            (70, 0.5196)
            (80, 0.5367)
            (90, 0.5117)
            (100, 0.5305)
            (110, 0.5364)
            (120, 0.5199)
            (130, 0.5412)
            (140, 0.5280)
            (150, 0.5341)
            (160, 0.5195)
            (170, 0.5161)
            (180, 0.5124)
            (190, 0.5361)
            (200, 0.5337)
            (210, 0.5331)
            (220, 0.5425)
            (230, 0.5385)
            (240, 0.5110)
            (250, 0.5204)
            (260, 0.5154)
            (270, 0.5036)
            (280, 0.5224)
            (290, 0.5250)
            (300, 0.5313)
            };
            \addlegendentry{L1}

        \addplot
            coordinates {
            (0, 0.5000)
            (10, 0.5000)
            (20, 0.5000)
            (30, 0.5000)
            (40, 0.5000)
            (50, 0.5000)
            (60, 0.5000)
            (70, 0.5000)
            (80, 0.5014)
            (90, 0.5228)
            (100, 0.5260)
            (110, 0.5302)
            (120, 0.5214)
            (130, 0.5200)
            (140, 0.5399)
            (150, 0.5524)
            (160, 0.5425)
            (170, 0.5366)
            (180, 0.5544)
            (190, 0.5481)
            (200, 0.5355)
            (210, 0.5539)
            (220, 0.5691)
            (230, 0.5580)
            (240, 0.5629)
            (250, 0.5482)
            (260, 0.5399)
            (270, 0.5169)
            (280, 0.5338)
            (290, 0.5496)
            (300, 0.5381)
            };
            \addlegendentry{L2}

        \addplot[mark=x]
            coordinates {
            (0, 0.5000)
            (10, 0.5000)
            (20, 0.5000)
            (30, 0.5000)
            (40, 0.5000)
            (50, 0.5000)
            (60, 0.5000)
            (70, 0.5000)
            (80, 0.5000)
            (90, 0.5135)
            (100, 0.5529)
            (110, 0.5866)
            (120, 0.5670)
            (130, 0.6007)
            (140, 0.5974)
            (150, 0.5925)
            (160, 0.6351)
            (170, 0.6205)
            (180, 0.6328)
            (190, 0.6621)
            (200, 0.6525)
            (210, 0.6652)
            (220, 0.6681)
            (230, 0.6574)
            (240, 0.6592)
            (250, 0.6911)
            (260, 0.6899)
            (270, 0.6764)
            (280, 0.6749)
            (290, 0.6596)
            (300, 0.6896)
            };
            \addlegendentry{L3}

            \draw[
                dashed
                ] 
                (axis cs:90, 0.5) -- (axis cs:90, 0.7);
        
        \end{axis}
    \end{tikzpicture}
    \caption{Layer-wise accuracy of the membership inference attack by Nasr et al.~\cite{nasr2019comprehensive} against intermediate models for the MNIST classification task. The model leaks more membership information as the number of training epochs grows. This behavior is particularly evident for the output layer.}
    \label{fig:attack_to_intermediate_model_memb_inf}
\end{figure}

Membership inference attacks aim to determine whether or not a given data point was part of the training set.
These attacks exploit the intrinsic difference in the model's behavior when performing prediction over known training data versus unseen data.
Membership inference attacks reveal how much a model retains from its training data, helping to gauge the potential effectiveness of other privacy attacks such as data reconstruction but can also pose significant privacy risks on their own.
Since the introduction of the first membership inference attack by Shokri et al.~\cite{shokri2017membership} in 2017, numerous studies have investigated the underlying causes of membership leakage in ML models.
The primary contributing factor to membership leakage appears to be model overfitting or poor generalization~\cite{shokri2017membership,yeom2018privacy}.
Several factors can exacerbate this issue, including a limited number of training samples~\cite{shokri2017membership,hilprecht2019monte}, high model complexity leading to overparametrization~\cite{nasr2019comprehensive}, and high feature dimensionality~\cite{shokri2017membership}.

\subsubsection*{Gray-Box Setting}
In the gray-box setting, the efficacy of membership inference attacks heavily depends on access to the last layers of the model.
While the initial layers of a neural network tend to extract simple features from the input, enabling them to generalize well, the later layers specialize in detecting higher-level abstract features in the input, making them prone to overfitting and memorizing the specific training examples.
For instance, in a CNN model trained for image classification, you can expect the first layers to learn more about edges and abstract shapes of the input image, while the last layers more about intricate texture and artifacts within those shapes~\cite{zeiler2014visualizing}.
Moreover, as the neural network progresses to the later layers, the parameter capacity increases, causing the target model to store information about the exact training samples~\cite{nasr2019comprehensive}.
Therefore, if the last layers of the model are accessible, membership inference attacks tend to be stronger due to the higher degree of membership information leakage.

\subsubsection*{Experimental Assessment}
To experimentally evaluate membership inference in the gray-box setting, we chose the white-box attack by Nasr et al.~\cite{nasr2019comprehensive} due to its component-wise approach.
Like other lines of work, their attack treats membership inference as a binary classification task, and trains a machine learning model to accomplish this task.
We employ the supervised version, in which the attacker is assumed to know a portion of the private training dataset and uses this knowledge to perform supervised training of the attack model.
Given a target data point, the attacker performs a feedforward pass of the model over it, computing hidden layer outputs, loss, and subsequent backpropagation to calculate gradients for each layer.
These computed values, along with the true label, serve as input features for the attack model.
While Nasr et al.'s work~\cite{nasr2019comprehensive} primarily focuses on the combination of the last layers in the model, our investigation aims to assess the privacy leakage of each individual layer of the model.

In line with the results by Nasr et al.~\cite{nasr2019comprehensive} attacking the last layers, we observe a similar effect for our generalized setting: the combination of multiple (intermediate) layers does not leak significantly more membership information than the just the last of those layers.
For instance, attacking layers 1, 3, and 4 of a given model does not provide a significant advantage over attacking just layer 4.
Consequently, we simplified our experimental setting and attack each layer individually and do not expect significantly different accuracy compared to attacks that include any combination of previous layers.

The number of members and non-members
is the same, resulting in a baseline attack accuracy of 50\%.
In \Cref{fig:memb_inf_eval}, we present the outcome of our experimental assessment on different datasets.
We report the average and maximum attack accuracy over four runs.
Our experimental results confirm the trend of the later layers of a model to leak more information compared to earlier layers.
In particular, the very last layer leaks considerably more information than the others.
This is especially evident in the cases of Purchase100 (\Cref{fig:memb_inf_eval:purchase100}) and MNIST (\Cref{fig:memb_inf_eval:MNIST}), where the attack accuracy for layer output increases from 55.12\% to 90.10\% (\mytilde 7.8 times increase offset to the baseline) and from 50.81\% to 64.49\% (\mytilde 17.8 times increase offset to the baseline), respectively, when passing from the second-to-last layer to the last layer.

Additionally, in line with the findings of Nasr et al.~\cite{nasr2019comprehensive}, our experiments confirm that the availability of gradients contributes to a higher attack accuracy.

\subsubsection*{Attacking Intermediate Models}
A model acquires more information about its training data the more training iterations it undergoes, thus leaking progressively more information as it approaches the end of the training process.
To assess how the membership leakage changes across the epochs, we use the attack by Nasr et al.~\cite{nasr2019comprehensive} against the intermediate models.
Specifically, in \Cref{fig:attack_to_intermediate_model_memb_inf}, we present the attack accuracy against the model trained on the MNIST dataset. We carry out the attack at 10-epoch intervals, targeting each layer within the model independently.
Our experiment reveals a consistent upward trend in attack accuracy with respect to the number of training epochs, especially for the later layers.
Notably, a significant deviation from the attack baseline appears only from epoch 90.

\subsection{Model Inversion}
\label{sec:attacks:model_inv}

\begin{figure}
    \centering
    \begin{tikzpicture}
        \draw (0, 0) node[inner sep=0, anchor=north west] {\includegraphics[width=0.95\columnwidth]{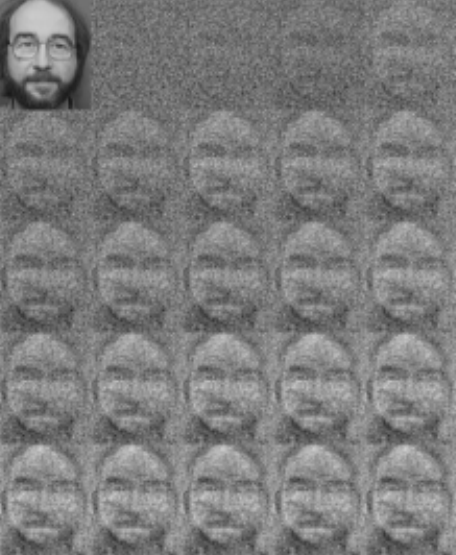}};
        \draw (1.61, 0.0) node[text=black, anchor=north west] {\small 0};
        \draw (3.22, 0.0) node[text=black, anchor=north west] {\small 1};
        \draw (4.83, 0.0) node[text=black, anchor=north west] {\small 2};
        \draw (6.44, 0.0) node[text=black, anchor=north west] {\small 3};
        \draw (0.00, -1.945) node[text=black, anchor=north west] {\small 4};
        \draw (1.61, -1.945) node[text=black, anchor=north west] {\small 5};
        \draw (3.22, -1.945) node[text=black, anchor=north west] {\small 6};
        \draw (4.83, -1.945) node[text=black, anchor=north west] {\small 7};
        \draw (6.44, -1.945) node[text=black, anchor=north west] {\small 8};
        \draw (0.00, -3.89) node[text=black, anchor=north west] {\small 9};
        \draw (1.61, -3.89) node[text=black, anchor=north west] {\small 10};
        \draw (3.22, -3.89) node[text=black, anchor=north west] {\small 11};
        \draw (4.83, -3.89) node[text=black, anchor=north west] {\small 12};
        \draw (6.44, -3.89) node[text=black, anchor=north west] {\small 13};
        \draw (0.00, -5.835) node[text=black, anchor=north west] {\small 14};
        \draw (1.61, -5.835) node[text=black, anchor=north west] {\small 20};
        \draw (3.22, -5.835) node[text=black, anchor=north west] {\small 30};
        \draw (4.83, -5.835) node[text=black, anchor=north west] {\small 40};
        \draw (6.44, -5.835) node[text=black, anchor=north west] {\small 50};
        \draw (0.00, -7.78) node[text=black, anchor=north west] {\small 60};
        \draw (1.61, -7.78) node[text=black, anchor=north west] {\small 70};
        \draw (3.22, -7.78) node[text=black, anchor=north west] {\small 80};
        \draw (4.83, -7.78) node[text=black, anchor=north west] {\small 90};
        \draw (6.44, -7.78) node[text=black, anchor=north west] {\small 100};
    \end{tikzpicture}
    \caption{Reconstruction of a face in the AT\&T dataset performed at different training epochs. The first picture is a class representative, while the number at the top-left of each picture denotes the corresponding training epoch of the model.}
    \label{fig:attack_to_intermediate_model_model_inv}
\end{figure}
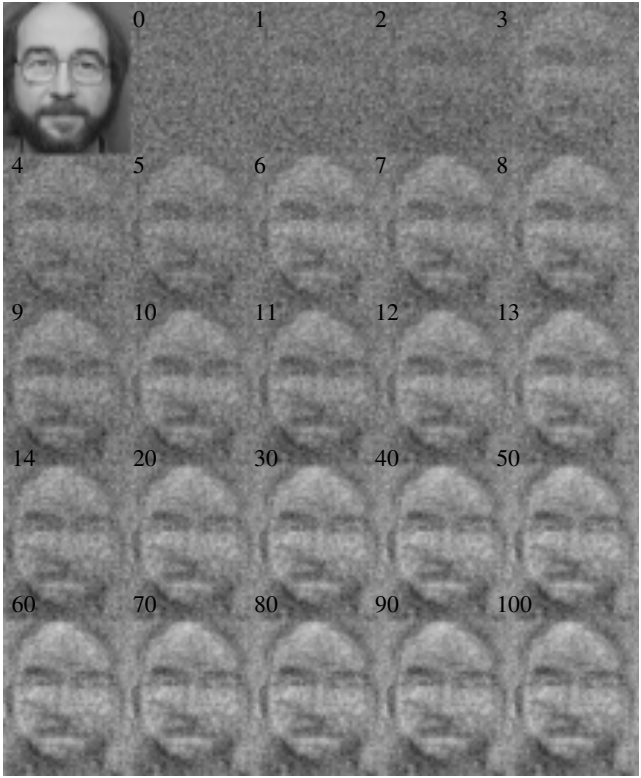

Model inversion, also known as reconstruction attack, aims to recreate training samples and, in some cases, their associated labels.
There are two main types of model inversion attacks: those that aim to reconstruct actual training samples~\cite{zhu2019deep,wang2019beyond} and those that aim to craft a class representative~\cite{fredrikson2015model,hitaj2017deep}.
The latter type is particularly useful when all samples associated with a given label are similar, such as faces of the same person, or when the attacker has no prior knowledge about what a specific label encodes.
The effectiveness of model inversion attacks has been shown to increase with the target model's level of overfitting~\cite{yeom2018privacy} and its predictive power, as measured by loss minimization~\cite{zhang2020secret}.
To mitigate these attacks, one suggested approach is to partially prune the gradients before updating the model~\cite{zhu2019deep}.

One of the first model inversion attacks on neural networks was developed by Fredrikson et al.~\cite{fredrikson2015model}.
The attacker crafts a dummy input for the target model and then uses gradient descent to optimize the dummy input.
The high-level idea is that, instead of fitting the model parameters to the input, the attacker computes the gradient of the loss function with respect to the input and fits the latter to the model parameters.
In contrast, other model inversion attacks use generative models to construct class representative. For instance, Hitaj et al.~\cite{hitaj2017deep} proposed a method based on Generative Adversarial Networks (GANs).
In this approach, the attacker designs a generator $G$ with the purpose of producing examples for a specific class $y$, using the target model itself as the discriminator.
The generator takes noise $x_\epsilon$ as input and generates $x_y = G(x_\epsilon)$, intended to represent class $y$.
The parameters of $G$, denoted as $\theta_G$, are optimized to minimize $L(\model(x_y),y)$, which indicates how confidently the model classifies $x_y$ as $y$.
Another model inversion attack by Zhu et al.~\cite{zhu2019deep} assumes that the attacker has access to the gradients $\nabla L(\model(x),y)$ computed by a training party for a data point $(x,y)$, e.g.,~in a collaborative learning setting when the attacker corrupts the supporting server (with no secure aggregation ongoing) or if the attacker corrupts all parties but the target one.
The attacker initializes a dummy data point $(x', y')$, computes the corresponding gradients $\nabla L(\model(x'),y')$, and minimizes the distance between these dummy gradients and the original gradients, which in turn brings the dummy input $(x',y')$ closer to the original $(x,y)$.
To solve the minimization problem, the attacker differentiates $\| \nabla L(\model(x'),y') - \nabla L(\model(x),y) \|$ with respect to $(x',y')$ and uses GD to find a local minimum $(x',y')$.

\subsubsection*{Gray-Box Setting}
All these attacks share the common requirement of computing the derivative of the target model's loss with respect to the model input: $\partial L / \partial x$, which can be written as $\frac{\partial L}{\partial x} = \frac{\partial L}{\partial l_1} \frac{\partial l_1}{\partial x}$, where $l_1$ is the output of the first layer.
To compute this derivative, the attacker then needs access to the first layer's gradients and parameters.

Blocking access to the first layer straightforwardly prevents the attack by Fredrikson et al.~\cite{fredrikson2015model}.
This limitation also prevents backpropagation from the target model to the generator for GAN-based approaches like\cite{hitaj2017deep}, and hinders the ability to solve the gradient difference minimization problem in the case of the attack by Zhu et al.~\cite{zhu2019deep}.
We conclude that denying the attacker access to the first layer of the model appears to be sufficient in preventing these specific types of model inversion attack, thus no experimental assessment is needed in this case.
However, we refrain from making a general claim, as there might still be potential attacks that can circumvent this limitation, and leave this as future research direction.

\subsubsection*{Attacking Intermediate Models}
We also evaluated the effectiveness of model inversion attacks against intermediate training models.
In \Cref{fig:attack_to_intermediate_model_model_inv}, we display the reconstruction of a face from the AT\&T dataset performed during different training epochs.
We use the model inversion attack by Fredrikson et al.~\cite{fredrikson2015model}.
Following their work, we train an MLP with one hidden layer with 3000 nodes, sigmoid activation function, and a softmax output layer, using the SGD optimizer and crossentropy loss function.
Similarly to membership inference, our experiments reveal that as the number of training epochs increases, the reconstructed class representative becomes more and more visually similar to the corresponding training examples.
However, we notice that the attack works already well even after just a few epochs.
This happens since a model inversion attack works well as soon as the model is generalizing well enough.

\subsection{Property Inference}
\label{sec:attacks:prop_inf}

Property inference attacks aim to extract properties about the training samples that are uncorrelated to the learning task at hand.
For example, in a face recognition task, where the goal is gender classification, the attacker might try to infer whether people in the training dataset are wearing sunglasses.
Similarly, for a model designed for handwriting recognition, the attacker may attempt to determine the font used to write the messages (e.g., cursive or block letters).

The underlying conditions and factors that enable property inference attacks are not yet fully understood~\cite{rigaki2020survey}.
It remains unclear what specific characteristics or vulnerabilities in a model make it susceptible to such attacks.
Surprisingly, these attacks have shown effectiveness even on well-generalized models, and the relationship between their efficacy and overfitting is still unclear~\cite{ganju2018property,melis2019exploiting}.
It has been suggested that sharing only a small portion of the gradients, as in the collaborative approach by Shokri and Shmatikov~\cite{shokri2015privacy}, may contribute to mitigate these attacks~\cite{melis2019exploiting}.

To carry out a property inference attack, the adversary needs access to samples both with and without the property they want to infer.
They calculate the gradients of the target model for both types of samples and trains a binary classifier to distinguish between the gradients of samples with the property of inference and samples without it.
In collaborative learning, the adversary can obtain the gradients of honest parties by computing the difference between two subsequent model updates.
However, if the adversary only has access to aggregated data from other parties, the attack may become less effective as the number of honest parties increases.

\subsubsection*{Gray-Box Setting}
Due to the lack of a clear understanding of the underlying causes of property inference leakage, it is challenging to predict how such attacks will scale in the gray-box setting.
The general idea is that the less gradients are exposed to the attacker, the less information is available for inference.
However, it remains unclear whether specific types of layers (e.g., convolutional or fully-connected) or their positions in the model contribute to higher or lower information leakage.

\subsubsection*{Experimental Assessment}
To assess this category of attacks in the gray-box setting, we build upon the white-box property inference attack proposed by Melis et al.~\cite{melis2019exploiting} and adapt it to target only a subset of the gradients.
This attack works in batches, aiming to determine whether a batch of data points possesses the target property or not.
In our variant of the attack, we feed to the attack model only the gradients computed with respect to parameters in the exposed layers.

We conduct the assessment on two datasets: Labeled Faces In the Wild (LFW) and EMNIST letters.
For the LFW dataset we train a CNN model following the architecture provided in~\cite{melis2019exploiting}, with 3 convolutional and 2 fully connected layers, using gender as main classification task, and race:black as inference task, which has been reported to yield the highest attack rate.
While for the EMNIST letter dataset, we train a custom MLP model with 3 layers, using the standard 26 letters classification as the main task, and the letter case (upper or lower) as the inference task.
The gradients of the exposed layers are fed to a Random Forest classifier with 50 trees, and the attack accuracy is averaged over multiple instances of the attack model.
For both datasets, we use batches of size 32, which are balanced with respect to the inference property, resulting in an attack accuracy baseline of 50\%.

For the CNN model trained on LFW, we did not find any specific patterns indicating whether some layers are more or less susceptible to inference than others.
The attack exhibited significant variability when conducting multiple iterations of training the target model and conducting repeated testing.
In \Cref{tab:prop_inf_LFW}, we report the attack accuracy layer-wise for multiple attempts of the experiment, revealing no consistent vulnerability or resistance of any layer across the runs.
On the other hand, for the MLP model trained on the EMNIST letters dataset, the attack did not achieve accuracy significantly above the baseline, regardless of the choice of the layers to attack (including combinations of multiple layers).

\begin{table}
    \caption{Property inference attack by Melis et al.~\cite{melis2019exploiting} against 5-layer CNN trained on the LFW dataset. The attack accuracy is reported per layer, demonstrating high variability across multiple training attempts of the same target model.}
    \begin{center}
        \begin{tabular}{|c|c|c|c|c|c|}
        \hline
        Run & Conv. 1 & Conv. 2 & Conv. 3 & Dense 1 & Dense 2 \\
        \hline
        1 & 66.88 & 81.44 & 88.13 & 95.06 & 81.69 \\
        2 & 60.25 & 74.31 & 72.00 & 53.19 & 56.13 \\
        3 & 63.31 & 69.63 & 79.06 & 66.25 & 55.88 \\
        4 & 86.69 & 93.44 & 91.75 & 99.88 & 96.00 \\
        \hline
        \end{tabular}
    \label{tab:prop_inf_LFW}
    \end{center}
\end{table}

\subsubsection*{Attacking Intermediate Models}
\begin{figure}
    \centering
    \begin{tikzpicture}
        \begin{axis}[
            legend columns=3,
            height=5cm,
            width=\columnwidth,
            xlabel={Training epoch},
            ylabel={Attack accuracy},
            enlarge x limits=0.05,
            enlarge y limits=0.05,
            legend style={at={(0.03,1.03)},anchor=south west},
            ymajorgrids=true,
        ]
        
        \addplot[color=blue,mark=*]
            coordinates {
            (0, 0.5000)
            (10, 0.5369)
            (20, 0.6181)
            (30, 0.5969)
            (40, 0.5756)
            (50, 0.6963)
            (60, 0.7350)
            (70, 0.7838)
            (80, 0.7638)
            (90, 0.6869)
            (100, 0.6981)
            };
            \addlegendentry{Conv. 1}

        \addplot[color=supergreen,mark=square*]
            coordinates {
            (0, 0.5000)
            (10, 0.7194)
            (20, 0.8356)
            (30, 0.8556)
            (40, 0.7038)
            (50, 0.8469)
            (60, 0.7138)
            (70, 0.7194)
            (80, 0.7388)
            (90, 0.7788)
            (100, 0.7963)
            };
            \addlegendentry{Conv. 2}

        \addplot[color=red,mark=otimes*]
            coordinates {
            (0, 0.5000)
            (10, 0.7475)
            (20, 0.8106)
            (30, 0.9044)
            (40, 0.7963)
            (50, 0.9038)
            (60, 0.7494)
            (70, 0.8313)
            (80, 0.7888)
            (90, 0.7963)
            (100, 0.7763)
            };
            \addlegendentry{Conv. 3}
        
        \addplot[color=purple,mark=triangle*]
            coordinates {
            (0, 0.5000)
            (10, 0.7538)
            (20, 0.7650)
            (30, 0.7906)
            (40, 0.7963)
            (50, 0.8800)
            (60, 0.6988)
            (70, 0.8006)
            (80, 0.9138)
            (90, 0.9500)
            (100, 0.9206)
            };
            \addlegendentry{Dense 1}

        \addplot[color=orange,mark=diamond*]
            coordinates {
            (0, 0.5000)
            (10, 0.6481)
            (20, 0.7450)
            (30, 0.6019)
            (40, 0.7350)
            (50, 0.8681)
            (60, 0.6600)
            (70, 0.7894)
            (80, 0.7988)
            (90, 0.8463)
            (100, 0.9200)
            };
            \addlegendentry{Dense 2}
        
        \end{axis}
    \end{tikzpicture}
    \caption{Layer-wise accuracy of a property inference attack against intermediate models for the LFW classification task.}
    \label{fig:attack_to_intermediate_model_prop_inf}
\end{figure}
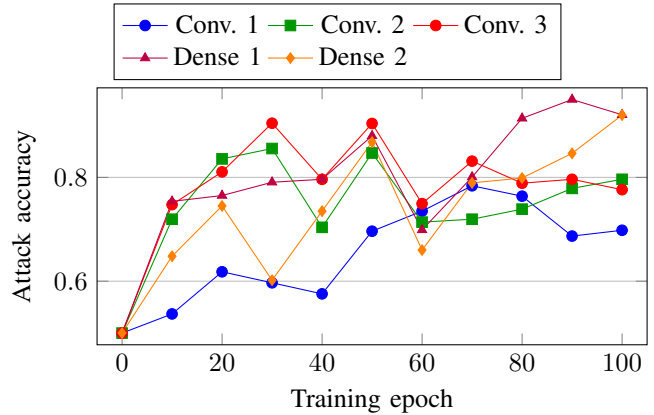

We also assessed how property inference varies across training epochs.
In \Cref{fig:attack_to_intermediate_model_prop_inf}, we report the attack accuracy of the property inference attack by Melis et al.~\cite{melis2019exploiting} against intermediate training models for the LFW classification task.
The attack is performed every 10 training epochs and it targets each layer of the model individually to get better insights.
The model exhibits a general upward trend for information leakage as the number of training epochs grows.
However, this trend appear not to be consistent, and the leakage is already substantial from the start.

\section{Partially Encrypted Models}
\label{sec:pems}

In this section, we leverage the insights from the investigation in \Cref{sec:attacks} to design a flexible collaborative learning protocol that allows users to trade off privacy for efficiency.
Our approach involves using an FHE scheme to encrypt the most vulnerable parts of the model and performing federated training on this partially encrypted model.
The level of privacy protection is determined by the selection of layers to be encrypted (\textit{secret layers} $\encryptedLayers$), while the remaining layers are left in plaintext (\textit{exposed layers} $\exposedLayers$).
The more layers we encrypt, the less information potential adversaries can access, thus enhancing privacy, but it also leads to more computations performed under encryption, thus reducing efficiency.
This flexibility allows our approach to achieve greater privacy than employing standard FL~\cite{mcmahan2017communication}, while achieving a more practical level of efficiency than fully encrypted solutions~\cite{froelicher2021scalable,sav2020poseidon}.

When performing feedforward and backpropagation on the model, we need to be careful about how to switch from secret to exposed layers and vice versa.
Computations are conducted under encryption whenever an encrypted layer is encountered, possibly invoking bootstrapping to refresh intermediate computations.
When an exposed layer is encountered, a decryption is called to allow continuing the training pass in plaintext (see \Cref{fig:partially_encrypted_nn}).

\subsection{Protocol Description}

We describe the protocol for MLP models, trained with SGD optimizer, and MSE loss, but it can be easily generalized to any feed-forward model.
Including simple momentum-based optimizers such as Nesterov Accelerated Gradient is straightforward and only requires an additional weight update.
While adaptive optimizers such as AdaGrad~\cite{duchi2011adaptive}, RMSProp~\cite{tieleman2012lecture}, and Adam~\cite{kingma2014adam} may require additional care due to the division by the rescaling coefficient.
For adapting the protocol to include convolutional layers, we refer to~\cite{sav2020poseidon}.

\subsubsection*{Global Training}
The protocol involves $\numParties$ training parties $\party_1, \dots, \party_\numParties$ and a central server $\centralServer$, whose role can potentially be taken by any $\party_i$.
The parties want to jointly train an MLP model, thus they agree on the model depth $\modelDepth$, architecture, activation functions, training hyper-parameters, and on the set of layers to encrypt $\encryptedLayers$.
The central server initiates the protocol by coordinating the FHE setup and key-generation phase, at the end of which $\party_1, \dots, \party_\numParties$ have their own private shares, and all actors possess the corresponding public key, relinearization, and rotation keys, which allow all parties to perform the necessary homomorphic operations.
The central server initializes the model $\model$ in plaintext, by generating random weight matrices $\weight_1, \dots, \weight_\modelDepth$ and bias vectors $\bias_1, \dots, \bias_\modelDepth$ of appropriate sizes, according to the distribution given by the chosen initialization technique.
Then, it uses the public key to encrypt the parameters $\weight_i, \bias_i$ corresponding to the secret layers $\layer_i \in \encryptedLayers$, after proper encoding (see \Cref{app:he_operations}).

At this point, the training starts, and proceeds as in standard FL.
The central server broadcast the partially encrypted model to $\party_1, \dots, \party_\numParties$.
Each party $\party_i$ performs a certain number of local training iterations, and sends back the updated local model to the central server.
The central server finally aggregates the local models, by averaging the parameters, using homomorphic addition and scalar multiplication by $1/\numParties$ for the ones in $\encryptedLayers$.
These steps are repeated for a fixed number of iterations $\numGlobalIt$ or until some convergence condition is satisfied (see Protocol~\ref{ptc:global_training}).

\begin{protocol}[htb]
    \caption{Global Training}
    \label{ptc:global_training}
    \begin{algorithmic}[1]
        \REQUIRE central server $\centralServer$, training parties $\party_1, \dots, \party_\numParties$
        
        \STATE Parties agree on
        model depth $\modelDepth$,
        architecture $\modelArch$,
        activation functions $\modelActFunc_j$,
        learning coefficient $\learningCoeff$,
        batch size $\localBatchSize$,
        parameter initialization technique $\modelInit$,
        number of global iterations $\numGlobalIt$,
        and the set of layers to encrypt $\encryptedLayers$%
        \\ FHE SCHEME SETUP:
        \STATE All actors collaborate to run the FHE setup and key generation, resulting in collective public key $\publicKey$%
        \\ MODEL INITIALIZATION:
        \STATE $\centralServer$ initializes the model in plaintext $(\weight_1, \bias_1), \dots, (\weight_\modelDepth,$ $\bias_\modelDepth) \gets \modelInit(\modelDepth, \modelArch)$
        \STATE $\centralServer$ encrypts parameters in secret layers:
        \FOR{$j = 1 \to \modelDepth$}
            \IF{$\layer_j \in \encryptedLayers$}
                \STATE $\weight_j \gets \encrypt(\publicKey, \weight_j)$
                \STATE $\bias_j \gets \encrypt(\publicKey, \bias_j)$
            \ENDIF
        \ENDFOR%
        \\ FEDERATED TRAINING
        \FORALL{$\globalIt = 1 \to \numGlobalIt$}
            \STATE $\centralServer$ sends $(\weight_1, \bias_1), \dots, (\weight_\modelDepth, \bias_\modelDepth)$ to the parties%
            \\ LOCAL TRAINING:
            \STATE Each $\party_i$ runs $\numLocalIt$ local model updates, getting the local model $(\weight_1^i, \bias_1^i), \dots, (\weight_\modelDepth^i, \bias_\modelDepth^i)$ and sends it to $\centralServer$ (see Protocol~\ref{ptc:local_training})%
            \\ AGGREGATION:
            \STATE $\centralServer$ aggregates the local models:
            \FOR{$j = 1 \to \modelDepth$}
                \STATE $\weight_j \gets \frac{1}{\numParties} \sum_{i = 1}^\numParties{\weight_j^i}$ \hfill (HE eval. if $\layer_j \in \encryptedLayers$)\ignorespaces
                \STATE $\bias_j \gets \frac{1}{\numParties} \sum_{i = 1}^\numParties{\bias_j^i}$ \hfill (HE eval. if $\layer_j \in \encryptedLayers$)\ignorespaces
            \ENDFOR%
        \ENDFOR%
    \end{algorithmic}
\end{protocol}

\subsubsection*{Local Training}
The local training subroutine, presented in Protocol~\ref{ptc:local_training}, involves each party $\party_i$ performing $\numLocalIt$ model updates locally before updating the central model.
During each local update, a batch of size $\localBatchSize$ is sampled from the local training set $\dataset_i$.
For each example in the batch, one training step is performed, which includes feedforward and backpropagation to compute gradients.
These gradients are then averaged across the batch sample and used to perform a local model update.
After $\numLocalIt$ updates, the local model is sent to the central server, which proceeds to the aggregation step.
We can add L2 regularization at the cost of an additional plaintext-scalar multiplication,
by multiplying the weight matrices by $1-\learningCoeff \decayCoeff / \localBatchSize$ just before line~\ref{ptc:local_training:line:weight_update}, where $\decayCoeff$ is the weight decay coefficient.

\begin{protocol}
    \caption{Local Training}
    \label{ptc:local_training}
    \begin{algorithmic}[1]
        \FOR{$\localIt = 1 \to \numLocalIt$}
            \STATE Sample $(x_1, y_1), \dots, (x_\localBatchSize, y_\localBatchSize)$ from local dataset
            \FOR{$\batchIndex = 1 \to \localBatchSize$}
                \STATE Compute gradients $(\nabla \weight_1^\batchIndex, \nabla \bias_1^\batchIndex), \dots, (\nabla \weight_\modelDepth^\batchIndex, \nabla \bias_\modelDepth^\batchIndex)$ by performing one training pass on $(x_\batchIndex, y_\batchIndex)$ (Protocol~\ref{ptc:one_train_pass})
            \ENDFOR
            \STATE Update local model:
            \FOR{$j = 1 \to \modelDepth$}
                \STATE $\weight_j \gets \weight_j - \frac{\learningCoeff}{\localBatchSize} \sum_{\batchIndex = 1}^\localBatchSize{\nabla \weight_j^\batchIndex}$ \hfill (HE eval. if $\layer_j \in \encryptedLayers$)\ignorespaces \label{ptc:local_training:line:weight_update}
                \STATE $\bias_j \gets \bias_j - \frac{\learningCoeff}{\localBatchSize} \sum_{\batchIndex = 1}^\localBatchSize{\nabla \bias_j^\batchIndex}$ \hfill (HE eval. if $\layer_j \in \encryptedLayers$)\ignorespaces
            \ENDFOR
        \ENDFOR
    \end{algorithmic}
\end{protocol}

Protocol~\ref{ptc:one_train_pass} outlines one training pass of our approach.
To feedforward an input through a partially encrypted model, we begin by feeding the vector in plaintext starting from the input layer.
When an encrypted layer is encountered, the computation proceeds under encryption, with bootstrapping being called when necessary (we omit bootstrapping calls from the protocol description since their call frequency depends on the FHE parameters).
As soon as an exposed layer is reached, a distributed decryption is invoked.
The same process is followed during the backpropagation step.
Note that if the last layer is encrypted, the loss is also computed under encryption.

\begin{protocol}
    \caption{One Training Pass}
    \label{ptc:one_train_pass}
    \begin{algorithmic}[1]
        \STATE $l_0 \gets x$%
        \\ FEEDFORWARD:
        \FOR{$j = 1 \to \modelDepth$}
            \STATE $u_j \gets l_{j - 1} \weight_j + \bias_j$ \hfill (HE eval. if $\layer_j \in \encryptedLayers$)\ignorespaces
            \IF{$j < \modelDepth \wedge \layer_j \in \encryptedLayers \wedge \, \layer_{j+1} \in \exposedLayers$}
                \STATE $u_j \gets \decrypt(u_j)$ \label{ptc:one_train_pass:line:output_decrypt}
            \ENDIF
            \STATE $l_j \gets \modelActFunc_j(u_j)$ \hfill (HE eval. if $(j = \modelDepth \wedge \layer_\modelDepth \in \encryptedLayers)$ \\ \hfill or $(j < \modelDepth \wedge \layer_j, \layer_{j + 1} \in \encryptedLayers)$)\ignorespaces
        \ENDFOR%
        \\ BACKPROPAGATION:
        \STATE $e_\modelDepth \gets y - l_\modelDepth$ \hfill (HE eval. if $\layer_\modelDepth \in \encryptedLayers$)\ignorespaces
        \STATE $\nabla \bias_\modelDepth \gets e_\modelDepth \modelActFunc'_\modelDepth(u_\modelDepth)$ \hfill (HE eval. if $\layer_\modelDepth \in \encryptedLayers$)\ignorespaces
        \STATE $\nabla \weight_\modelDepth \gets \nabla \bias_\modelDepth l_{\modelDepth-1}^T$ \hfill (HE eval. if $\layer_\modelDepth \in \encryptedLayers$)\ignorespaces
        \FOR{$j = \modelDepth - 1 \to 1$}
            \STATE $e_j \gets \nabla \bias_{j+1} \weight_{j+1}^T$ \hfill (HE eval. if $\layer_{j+1} \in \encryptedLayers$)\ignorespaces
            \IF{$\layer_j \in \exposedLayers \wedge \, \layer_{j+1} \in \encryptedLayers$}
                \STATE $e_j \gets \decrypt(e_j)$ \label{ptc:one_train_pass:line:error_decrypt}
            \ENDIF
            \STATE $\nabla \bias_j \gets e_j \modelActFunc'_j(u_j)$ \hfill (HE eval. if $\layer_j, \layer_{j+1} \in \encryptedLayers$)\ignorespaces
            \STATE $\nabla \weight_j \gets \nabla \bias_j l_{j-1}^T$ \hfill (HE eval. if $\layer_j, \layer_{j-1} \in \encryptedLayers$ \\ \hfill or $\layer_j, \layer_{j+1} \in \encryptedLayers$)\ignorespaces
        \ENDFOR%
    \end{algorithmic}
\end{protocol}

Unless we are at the last layer of the model, decrypting just after the linear transformation at line~\ref{ptc:one_train_pass:line:output_decrypt} is optimal.
To show this, let us consider the case we are at the end of a group of encrypted layers, that is we are at layer $\layer_j$ for some $j < \modelDepth$, with $\layer_j \in \encryptedLayers$ and $\layer_{j+1} \in \exposedLayers$.
If instead of decrypting $u_j$, we perform an additional step under encryption, and decrypt after the evaluation of $\modelActFunc_j(u_j)$, then an adversary could just invert the activation function if bijective (e.g., sigmoid) or still get information about $u_j$ for most of the activation functions commonly used.
If we keep going under encryption for a step further, and decrypt for instance after the next linear transformation $u_{j+1}$ in order to keep $l_j$ private, then, since $\layer_{j+1} \in \exposedLayers$, we would need to invoke a decryption for $\nabla \weight_{j+1}$, which depends on $l_j$. However, an adversary could easily retrieve $l_{j}$ given $\nabla \weight_{j+1}$ and $\nabla \bias_{j+1}$, which is also in plaintext since $\layer_{j+1} \in \exposedLayers$.
Similar remarks hold for the decryption of the error in the backpropagation phase at line~\ref{ptc:one_train_pass:line:error_decrypt}.

Consequently, when encrypting a single non-output layer, the corresponding layer output and gradient cannot be protected.
Thus, to protect the initial or central portions of the model, at least two consecutive layers need to be encrypted, and even in that case, the gradient of the bias of the last layer of such group will be exposed.
One approach to address this limitation is to omit the bias parameters on that specific layer.

An additional argument against encrypting only one non-output layer is the potential for an adversary to reconstruct the encrypted parameters.
If the adversary can gather enough input and output pairs $(x, y)$, where $y = \weight x + \bias$, they could retrieve the values of $\weight$ and $\bias$ by solving a system of linear equations with the layer parameters as the unknowns.
Encrypting two consecutive layers (or the last one), already makes the system of equation significantly harder to solve.
The system will involve many more variables and the activation function of the first layer as well, which is typically non-linear.
As additional measure, lowering the precision of the FHE scheme and introducing additional noise in the ciphertext can further complicate the reconstruction of the encrypted layers.

\subsubsection*{Prediction}
After completing the training phase, the partially encrypted model can be directly used for predictions.
However, cooperation among the training parties remains necessary for distributed bootstrapping and decryption calls.
Prediction queries can be initiated by any of the training parties or an external entity.
If the querier is one of the training parties, they can locally conduct the feedforward step and seek assistance from the others only for bootstrapping and decryption, including the potential output decryption if the last layer is also encrypted.
If the querier is an external entity, additional precautions are needed due to the presence of exposed layers.
To ensure the privacy of the query, the querier encrypts their input with the collective public key of the FHE scheme, and sends the encrypted input to one of the training parties.
The selected party will then perform the feedforward pass on behalf of the querier.
In this case, operations on the exposed layers must be adapted to work under encryption.
While matrix multiplication and bias addition remain straightforward, the activation functions need to be approximated to be homomorphically evaluated, leading to a potential loss of accuracy.
Moreover, unlike during training, the output of a group of adjacent layers should not be decrypted.
\footnote{It is possible to perform plaintext operations in exposed central layers, when they are sufficiently distant from both the input and output layers to lower the possibility of reconstructing the query input or output.}
The final output of the model then remains encrypted, regardless of whether the last layer is private or not.
At this point, the training parties can send the decryption shares of the output to the querier, who can then reconstruct the output in clear, or a \textit{key-switch}~\cite{mouchet2021multiparty} to the public key of the querier can be performed.

\subsubsection*{Delayed Encryption}
We can leverage the investigation of privacy attacks on intermediate training models to optimize our solution.
From \Cref{sec:attacks}, we know that some attacks start to be effective only after some number of training epochs.
An optimization for our solution consists of starting the federated learning process fully in plaintext, and encrypting (some of) the layers only once the model becomes vulnerable.
The number of layers to encrypt can be dynamically adjusted during the training process.
We will discuss this more in details in \Cref{sec:pems:procedure}.

\subsection{Efficiency Analysis}
\label{sec:pems:efficiency}

Compared to a fully encrypted approach, partially encrypted models offer significant efficiency advantages, including reduced number of computations under encryption, lighter model updates, and fewer communication rounds.
To simplify the analysis, we assume all layers to have same order of magnitude sizes, and plaintext size and operation costs to be negligible with respect to their encrypted counterpart.

In general, since computations are carried out in plaintext in the exposed layers, we expect a lower-bound for the gain in computational complexity to be at least linear in the number of exposed layers relative to the total number of layers in the model, i.e. $|\exposedLayers| / \modelDepth$.
Additionally, we can avoid the homomorphic evaluation of the last activation function in each group of encrypted layers (but the ones containing the last layer).
Moreover, as the training parties decrypt their computations at the end of each encrypted block, the need for distributed bootstrapping decreases or even disappears.
This results in gains in computation efficiency, communication size, and communication rounds, as each bootstrapping process typically requires one round-trip of communication.
This advantage is amplified by the fact that bootstrapping in the CKKS scheme needs the ciphertext to have some levels left, further increasing the computational overhead in fully encrypted models.
Depending on the number of contiguous encrypted layers, even faster somewhat HE schemes can be adopted. 

Regarding the communication size during model update and broadcast, we again observe a linear gain in $|\exposedLayers| / \modelDepth$, as the model parameters corresponding to exposed layers are sent in plaintext.
On the other hand, during the aggregation phase, fewer parameters need to be averaged under encryption, leading to an additional gain in terms of computational complexity.
Finally, we note that in the optimized version of our solution, the performance gain factor $|\exposedLayers| / \modelDepth$ changes according to the number of exposed layers across the training epochs.
We provide detailed execution time and communication size of a specific use case in \Cref{sec:pems:exp_evaluation}.

\subsection{Security Analysis}
\label{sec:pems:security}

In this section, we sketch a security proof for the encrypted layers of the model, and we discuss the capabilities of different types of adversaries for each class of privacy attack.

Our approach aims to preserve the privacy of the training data, during both the training and prediction phases, by encrypting the most vulnerable layers of the model.
In the semi-honest setting, we prove that no party, including the central server, can learn more information about the training data of any other party or the model parameters corresponding to any layer in $\encryptedLayers$, other than what can be deduced from their own data (including the model output, in case of predictions), and from the parameters and intermediate computations of the layers in $\exposedLayers$.
In the case of predictions requested by an external entity, we can make this claim stronger, as the querier should not learn anything, other than what can be deduced from only their own input data and the query output.

For the security proof, we proceed as in~\cite{sav2020poseidon}, but assuming the simulator is also given access to the parameters of the exposed layers at each iteration, and to the output of each decryption call.
The idea is to see the overall scheme as a composition of the underlying FHE protocols, which are all simulatable.
For the basic protocols like key generation and decryption, we rely on the proofs by Mouchet et al.~\cite{mouchet2021multiparty}, while for the distributed bootstrapping, we rely on the proof by Sav et al.~\cite{sav2020poseidon}.
Note that the security of our approach does not hold for a malicious adversary, which can exploit the decryption call in the protocol decrypt the secret layers.

In the rest of the section, we discuss whether specific attacks can still run on the exposed layers.
We consider three threat model configurations, based on the possible combinations of the ML and cryptographic adversary's capabilities introduced in \Cref{sec:threat_model}:
\begin{enumerate}
    \item ML-passive and crypto-passive, where the adversary follows the protocol and can only use inputs from the original dataset (no maliciously crafted inputs);
    \item ML-active and crypto-passive, where the adversary follows the protocol but may craft malicious inputs for the training procedure;
    \item ML-active and crypto-active, where the adversary can arbitrarily deviate from the protocol and may craft arbitrary malicious input.
\end{enumerate}
Note that the distinction between the first two settings is important in real-world scenarios, since it has implications in terms of detectability and liability.
If a client becomes corrupted during the training process, an ML-active attack is potentially more detectable than an ML-passive attack.
Detection can occur by analyzing the intermediate updates or the final model, or by employing some form of commitment to the training dataset.

In \Cref{tab:discussing_attacks}, we outline the capabilities of each attack discussed in \Cref{sec:attacks} for the threat model configurations described above.
Note that, in contrast to FL in plaintext, the presence of encrypted layers restricts the adversary from freely conducting any inference on the model.

\begin{table*}[htbp]
    \caption{Description of attack capabilities in different threat models, for different privacy attacks. For a description of the active variant of the attacks, we refer to the corresponding works.}
    \begin{center}
        \resizebox{\textwidth}{!}{%
        \begin{tabular}{p{28pt}p{22pt}p{127pt}p{157pt}p{130pt}}
        \hline
        \textbf{Attack} & \textbf{Class} & \textbf{ML-passive \& crypto-passive} & \textbf{ML-active \& crypto-passive} & \textbf{ML-active \& crypto-active} \\
        \hline
        Nasr \newline et al.~\cite{nasr2019comprehensive} & Memb. \newline inference & The attack is not possible, since the attacker cannot pass its target data point through the model. & The attack is possible, but limited to its passive variant on the exposed layers. & Also the active variant of the attack is possible, since the attacker can perform gradient ascent. \\[20pt]
        \hline
        Fredrikson \newline et al.~\cite{fredrikson2015model} & Model \newline inversion & The attack is not possible, since the attacker cannot pass the dummy input through the model. & The attack is not possible if the first layer is encrypted, since the attacker cannot backpropagate over the input layer. & The attack is always possible, the malicious adversary can decrypt the first layer if necessary. \\[20pt]
        \hline
        Hitaj \newline et al.~\cite{hitaj2017deep} & Model \newline inversion & The attack is not possible, since the attacker cannot pass the generator output through the model. & The attack is not possible if the first layer is encrypted, since the attacker cannot backpropagate to the generator. & The attack is always possible, the malicious adversary can decrypt the first layer if necessary. \\[20pt]
        \hline
        Zhu \newline et al.~\cite{zhu2019deep} & Model \newline inversion & The attack is not possible, since the attacker cannot pass the dummy input through the model. & The attack is not possible, since the attacker cannot compute the derivative of the gradients with respect to the input. & The attack is always possible, the malicious adversary can decrypt any layer if necessary. \\[20pt]
        \hline
        Melis \newline et al.~\cite{melis2019exploiting} & Property \newline inference & The attack is not possible, since the attacker cannot pass the target batch through the model. & The attack is possible, but limited to the exposed layers. The active variant is not possible if the last layer is encrypted. & The attack is always possible, the malicious adversary can decrypt any layer if necessary. \\
        \hline
        \end{tabular}
        \label{tab:discussing_attacks}
        }
    \end{center}
\end{table*}

\subsection{Differential Privacy and the Protection of Exposed Layers}

We discuss the relation between our work and solutions based on differential privacy, and how the latter can be used to protect the exposed layers in our approach.
The trade-offs offered by our solution and by DP are substantially different: privacy against efficiency and privacy against accuracy, respectively.
In particular, our solution does not compromise the model's accuracy, hence providing more utility than DP, while at the same time being more efficient than fully encrypted approaches, posing our work in between FHE-based and DP-based solutions.

DP can also be incorporated in our approach and used to mitigate the leakage from the exposed layers, providing in this way theoretical guarantees for our framework.
We describe a possible implementation in \Cref{app:dp}, where we show how to adapt the approach by Shokri and Shmatikov~\cite{shokri2015privacy} to partially encrypted models.
You can see our solution as a way to reduce the consumption of the privacy budget by encrypting some of the layers.
The save in terms of privacy budget is proportional to the number of parameters that are encrypted over the total amount of parameters of the model.

\subsection[Procedure to Choose Layers to Encrypt]{Procedure to Choose $\encryptedLayers$}
\label{sec:pems:procedure}

Since the privacy leakage of a model strongly depends on the training dataset, there is no general-purpose guideline on how many and which layers to encrypt.
A practical approach we propose consists of getting a lower bound estimate on the privacy leakage through an assessment on the local training data.
To do so, the parties can train a dummy model on their own private datasets and perform a privacy assessment locally.
Instead of exploring all possible combinations of private-exposed layers, the parties can rely on the insights discussed in \Cref{sec:attacks} to determine which configurations are the most meaningful to assess.
Since each local dataset is a subset of the joint dataset, the privacy leakage assessed locally provides an empirical worst-case for the privacy leakage of the joint model.
From the efficiency point of view, the parties can use the insights from \Cref{sec:pems:efficiency}, by also taking into account their specific computation and communication constraints (e.g., bandwidth and network delay between the nodes).
Finally, the parties can collectively agree on which layers to encrypt by leveraging MPC techniques, avoiding leaking potential information about their local dataset.
Depending on the specific situation and requirements, the parties can perform a majority vote or compute the union of the local choices to reach a consensus on the layers to be encrypted.

We provide an example of how to agree on $\encryptedLayers$ for the optimized version of our solution, in case membership inference attacks are considered.
Each party $\party_i$ trains a model locally and assess it against the considered privacy attack across the training epochs, as in \Cref{fig:attack_to_intermediate_model_memb_inf}.
Then, they select a privacy threshold $\privacyThreshold_i \in [0.5, 1]$, which fixes an upper bound on the model leakage they are willing to allow.
The party then proceeds to compute their preferred choice for the layers to encrypt $\encryptedLayers^{i,\globalIt}$ for each epoch $\globalIt = 1, \dots, \numGlobalIt$ as the minimum set of layers that keeps the attack accuracy under $\privacyThreshold_i$.
For consistency reasons, if a layer $\layer_j$ is included in $\encryptedLayers^{i,\globalIt}$, then all subsequent layers $\layer_k$ for $k > j$ should be included as well.
Moreover, the layer should be included in all the future epochs as well, that is $\layer_j \in \encryptedLayers^{i,\globalIt'}$ for all $\globalIt' > \globalIt$.
Then, the parties use MPC to compute $\encryptedLayers^\globalIt$ as the union of the $\encryptedLayers^{i,\globalIt}$ for each $\globalIt = 1, \dots, \numGlobalIt$.

\subsection{Experimental Evaluation}
\label{sec:pems:exp_evaluation}

In this section, we experimentally evaluate our partially encrypted model approach for different choices of $\encryptedLayers$.
Our experiments show a similar run-time and communication performance of our framework for all datasets mentioned in \Cref{sec:attacks:memb_inf}. For space limitations, we report only the results for MNIST.
The performance for a general-architecture MLP is discussed in \Cref{sec:pems:efficiency}, and can be extrapolated using the microbenchmark in \Cref{tab:pem_microbenchmark}. While for all datasets, the privacy evaluation is reported in \Cref{sec:attacks}.

\subsubsection{Experimental Setup}
We implement \tool\footnote{\repository} in C++, building on top of the OpenFHE library\footnote{\url{https://github.com/openfheorg/openfhe-development}} for the multiparty CKKS functionalities.
Our implementation uses CKKS with a 5-bit integral precision, 55-bit decimal precision (scaling factor), a moduli tower with 8 levels, and a cyclotomic ring degree of $2^{15}$.
To assess the performance of our prototype in a realistic scenario, we run the experiments within Mininet\footnote{\url{https://github.com/mininet/mininet}}, a network emulator that allows us to configure different network topologies and impose constraints on bandwidth and network delay.
Different virtual hosts are spawned within a server with an Intel Xeon Platinum 8358 running at 2.60 GHz, with 64 threads on 32 cores, and 512 GB RAM.
In particular, for the evaluation, we consider a setup with 3 training parties and a central server, communicating over TCP in a star topology network.
The communication is constrained by 1Gbps bandwidth and 10ms network delay between the nodes.

Each party is provided with 30 examples from the MNIST dataset, and they jointly train an MLP model with two hidden layers of size 30, 20.
Each layer uses sigmoid activation functions, which is approximated in $[-10,10]$ by a polynomial of degree 13 for HE evaluation.
To simplify the analysis of the trade-off given by the choice of $\encryptedLayers$,
we decided to focus on the specific case of encrypting only one group of contiguous layers containing the output layer, which, from the investigation in \Cref{sec:attacks}, seems to be one of the most meaningful settings for our approach when considering inference attacks.
That is, given a model $\model$ of depth $\modelDepth$, we then have $\exposedLayers = \left\{\layer_1, \dots, \layer_\layerThreshold\right\}$ and $\encryptedLayers = \left\{\layer_{\layerThreshold + 1}, \dots, \layer_\modelDepth\right\}$ for some $\layerThreshold \in \left\{0, \dots, \modelDepth\right\}$.
This way, the trade-off is controlled by the one-dimensional parameter $\layerThreshold$: when $\layerThreshold = 0$ we are in the extreme case of FL with full encryption of the model~\cite{sav2020poseidon}, while when $\layerThreshold = \modelDepth$ we are in the extreme of FL in plaintext~\cite{mcmahan2017communication}.
We report the results for the non-optimized version of our framework.

\subsubsection{Performance Results}
As shown in \Cref{tab:pem_performance_MNIST}, the efficiency of our approach scales approximately linearly with $\layerThreshold$, both in terms of run time and communication size.
\begin{table}
    \caption{Execution time and communication size of our approach on the MNIST dataset for a 3-layer MLP, with varying levels of layer encryption: none ($\layerThreshold = 3$), last layer ($\layerThreshold = 2$), last two layers ($\layerThreshold = 1$), and full model encryption ($\layerThreshold = 0$).}
    \begin{center}
        \begin{tabular}{|c|r|r|r|r|}
        \hline
        $\layerThreshold$ & \multicolumn{2}{c|}{Training} & \multicolumn{2}{c|}{Inference} \\
        \cline{2-5}
        & \multicolumn{1}{c|}{Run time} & \multicolumn{1}{c|}{Comm. size} & \multicolumn{1}{c|}{Run time} & \multicolumn{1}{c|}{Comm. size} \\[-2pt]
        & \multicolumn{1}{c|}{(h)} & \multicolumn{1}{c|}{(GB)} & \multicolumn{1}{c|}{(s)} & \multicolumn{1}{c|}{(MB)} \\
        \hline
        3 &  1.8e-2 & 3.9e-2 & 5.0e-3 & 0.0 \\
        2 &  726.7 &  980.8 & 23.5 & 10.3 \\
        1 & 1504.1 & 1988.8 & 50.0 & 23.3 \\
        0 & 2232.7 & 3083.4 & 76.8 & 36.3 \\
        \hline
        \end{tabular}
    \label{tab:pem_performance_MNIST}
    \end{center}
\end{table}
The reported run time and communication size have been averaged among the training parties for consistency. In particular, the communication size refers to the total volume of messages received and sent per party over 300 epochs of training.
The model achieves the same test accuracy as its plaintext counterpart (i.e., 48.7\%). The noise from the FHE scheme and the approximation error of the activation functions do not have a significant impact on the training procedure.

\begin{figure}[htb]
    \centering
    \begin{tikzpicture}
        \pgfplotsset{
            selective show sum on top/.style={
                /pgfplots/scatter/@post marker code/.append code={%
                    \ifnum\coordindex=#1
                       \node[
                       at={(normalized axis cs:%
                           \pgfkeysvalueof{/data point/x},%
                           \pgfkeysvalueof{/data point/y})%
                       },
                       anchor=south,
                       ]
                       {\pgfmathprintnumber{\pgfkeysvalueof{/data point/y}}};
                    \fi
                },
            },selective show sum on top/.default=0
        }
        \begin{axis}[%
                    ybar stacked,
                    ymin=0,
                    bar width=24pt,
                    symbolic x coords={3, 2, 1, 0},
                    xtick=data,
                    nodes near coords={
                        \ifnum\coordindex=3\else
                            \pgfmathprintnumber\pgfplotspointmeta
                        \fi
                    },
                    ylabel=Run time (s),
                    xlabel=$\layerThreshold$,
                    major x tick style = transparent,
                    enlarge x limits=0.15,
                    legend cell align={left},
                    legend pos=north west,
                    legend style={
                        column sep=1ex
                    }
                ]
            \addplot[fill=blue!60,font=\scriptsize] coordinates {
                (3,0.0)
                (2,46.8)
                (1,84.8)
                (0,134.9)
            };
            \addplot[fill=red!60,selective show sum on top/.list={0,1,2,3},font=\scriptsize] coordinates {
                (3,0.0)
                (2,236.8)
                (1,477.4)
                (0,739.3)
            };
            \node[font=\scriptsize,anchor=south] at (3,0) {0.0};
            \legend{Computation,Communication}
        \end{axis}
    \end{tikzpicture}
    \caption{Computation vs. communication time for one training pass in our approach on MNIST for a 3-layer MLP, with varying levels of encryption: none ($\layerThreshold = 3$), last layer ($\layerThreshold = 2$), last two layers ($\layerThreshold = 1$), and full model encryption ($\layerThreshold = 0$).}
    \label{fig:pem_performance_MNIST_comp_vs_comm_time}
\end{figure}
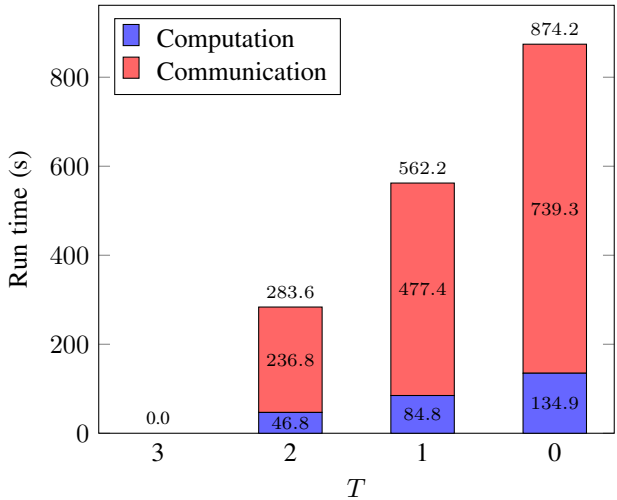

In \Cref{fig:pem_performance_MNIST_comp_vs_comm_time}, we observe that communication time is the dominant factor on the overall performance.
Note that the high communication and overall run time are partly due to our current implementation being still a prototype, and the use of relatively high-degree approximation for the activation functions.
We acknowledge that there are large margins for optimization, starting from compressing the model updates before transmitting them.
Also, using a tile-packing of the weight matrices as suggested in~\cite{aharoni2023helayers} may increase the performance of our approach.
However, our main focus is on comparing the relative efficiency measures between different choices of encrypted layers, rather than their absolute values.

In \Cref{app:microbenchmark}, we present the micro-benchmarks for most HE functionalities used in our approach.

\subsubsection{Privacy-Efficiency Trade-Off}
The more layers we encrypt, the higher the privacy, as less parameters are available to a potential adversary.
However, encrypting more layers also leads to lower performance due to the overhead introduced by homomorphic evaluations and distributed bootstrapping.
Thus, there is a trade-off between privacy and efficiency when deciding how many and which layers to encrypt in a model.
The optimal choice depends on the specific use case, in particular on the types of attacks one wants to protect the model from, and the desired balance between privacy and performance.

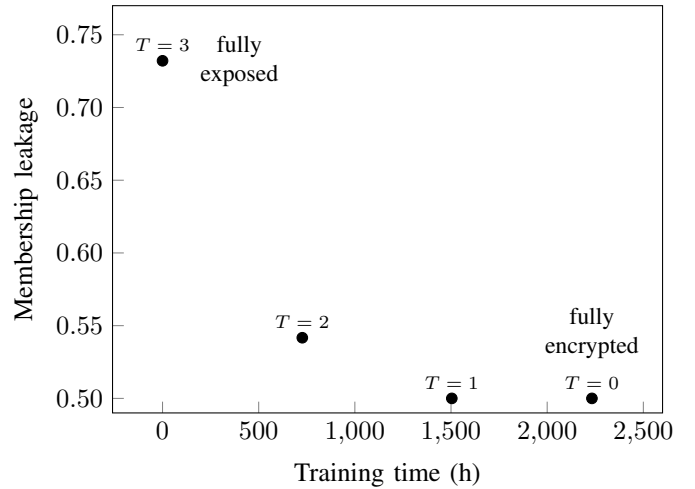
\begin{figure}[htbp]
\centering
    \begin{tikzpicture}
        \begin{axis}[
                height=7cm,
                width=\columnwidth,
                scatter,          
                only marks,       
                xlabel=Training time (h),
                ylabel=Membership leakage,
                xtick pos=bottom,
                ytick pos=left,
                xmax=2600,
                ymin=0.49,
                ymax=0.77,
                nodes near coords,
                point meta=explicit symbolic, 
                yticklabel style={
                    /pgf/number format/.cd,
                    /pgf/number format/fixed,
                    /pgf/number format/fixed zerofill,
                    /pgf/number format/precision=2,
                    }
            ]
            \addplot[mark options={black},font=\scriptsize] table[meta=label] {
                x   y       label
                2232.7 0.5000  {$\layerThreshold = 0$}
                1504.1 0.5000  {$\layerThreshold = 1$}
                726.7 0.5417  {$\layerThreshold = 2$}
                0.017 0.7321  {$\layerThreshold = 3$}
            };
    
            \node[text width=40pt,align=center] at (axis cs: 400,0.7321) {\small fully exposed};
            \node[text width=40pt,align=center] at (axis cs: 2232.7,0.545) {\small fully encrypted};
        \end{axis}
    \end{tikzpicture}
\caption{Membership inference~\cite{nasr2019comprehensive} on gradients, assuming attacker corrupted 2 out of 3 parties.}
\label{fig:trade_off}
\end{figure}

In \Cref{fig:trade_off}, we represent this trade-off, using membership inference accuracy as the metric for privacy leakage.
Each point on the plot represents a particular configuration of our approach, for different choices of $\encryptedLayers$.
The optimum point of the trade-off occurs when both the leakage and the training time are minimized (i.e., at the origin point of the chart).

\subsubsection{Comparison with Prior Work}
The flexibility of our approach allows the user to sacrifice some privacy in order to gain training performance in terms of computation and communication time compared to fully encrypted solution like SPINDLE~\cite{froelicher2021scalable} or POSEIDON~\cite{sav2020poseidon}.
At the same time, it provides higher privacy levels than training entirely in plaintext, without compromising significant accuracy.
Note that setting $\layerThreshold = 0$ corresponds to the original POSEIDON idea resulting in a fully encrypted model.
\footnote{POSEIDON is not public accessible, hence we cannot directly compare their run time measures with ours. All potential optimizations are directly compatible with our construction and vice versa all our optimizations are compatible with POSEIDON.}

For the very specific case presented in \Cref{fig:trade_off}, encrypting only the last layer ($\layerThreshold = 2$) provides a good trade-off.
It offers a membership leakage very close to a random guess (54.17\%, a 5.6 times reduction from the fully plaintext solution's 73.21\%, relative to the random guess baseline), while reducing the training time with respect to the fully encrypted solution by a factor of 3.1.
Assuming the adversary does not possess the target label, the leakage reduction factor increases to 17.8.
While using the optimized version of our solution increases the run-time gain factor to 4.03 (by encrypting no layers till epoch 90, and only the last layer afterwards).
The advantage provided by our solution may become even more evident for models with deeper architectures~\cite{krizhevsky2012imagenet,agarap2019training,dosovitskiy2020image}, particularly in settings with constrained communication networks.

\section{Related Work}
\label{sec:rel_work}

In addition to our overview of the literature on privacy attacks in~\Cref{sec:attacks}, we provide a brief overview of related work on PPML in this section.
In general, the PPML literature can be categorized into works that focus on ensuring privacy against adversaries during either the inference or training stage.

\subsection{Privacy-Preserving Inference}
In this scenario, an already trained model is typically sent to a cloud server, which provides predictions as a service on behalf of the model's owner.
The main goal is to ensure the confidentiality of the user's query input and the corresponding output.
This kind of oblivious prediction functionality can be achieved in multiple ways.
One approach involves using leveled Homomorphic Encryption to encrypt the query input and perform the inference homomorphically (CryptoNets~\cite{gilad2016cryptonets}).
Another method involves using MPC among a cluster of non-colluding servers, possibly mixed with Garbled Circuits and HE (Gazelle~\cite{juvekar2018gazelle}, MiniONN~\cite{liu2017oblivious}, Chameleon~\cite{riazi2018chameleon}, CryptFlow~\cite{kumar2020cryptflow}).
Some of those MPC solutions, by distributing the model across multiple servers, prevent the cloud servers from accessing or stealing the model. This aspect becomes particularly valuable when the model owner lacks trust in the service provider.
Some line of work focuses on exploring inference on ML models in Trusted Execution Environments (TEEs), where the main challenges are ensuring performance and resiliency against side-channels attacks (Slalom~\cite{tramer2018slalom}).

In addition, to protect against users attempting to retrieve information about the training data or reconstruct the model through a smart choice of queries (i.e., black-box attacks), various defense mechanisms have been proposed.
General purpose solutions like using DP during model training (Song et al.~\cite{song2013stochastic}, Abadi et al.~\cite{abadi2016deep}) have been widely studied,
as well as specific defense strategies against certain attacks.
For instance, possible strategies against inference attacks include
perturbing the prediction vector with noise (Memguard~\cite{jia2019memguard}),
masking the output confidence score by only revealing the top-k scores or just the prediction label (Shokri et al.~\cite{shokri2017membership}, Choquette-Choo et al.~\cite{choquette2021label}, Li et al.~\cite{li2021membership})
employing adversarial regularization by jointly minimizing the classification loss and maximizing a theoretical attack model’s loss (Nasr et al.~\cite{nasr2018machine}),
using knowledge distillation to put distance between the private dataset and the deployed model (PATE~\cite{papernot2016semi}, Shejwalkar et al.~\cite{shejwalkar2021membership}),
or simply applying standard regularization techniques (Shokri et al.~\cite{shokri2017membership}).
Furthermore, other approaches have been developed to prevent model stealing, such as PRADA~\cite{juuti2019prada}, which checks for structured patterns in user queries, or like PrivDNN~\cite{ren2024privdnn}, which encrypts the individual neurons that contribute the most to the model's utility. It is important to note, however, that while the last two approaches enhance model privacy, they do not inherently safeguard the confidentiality of the training data.

\subsection{Privacy-Preserving Training}
To introduce adversaries during the training stage, we need to consider a collaborative learning setting.
In this setting, adversaries may corrupt one or more training clients and potentially the supporting server, making it possible to run privacy attacks on the intermediate model updates.
Differential privacy can be employed to inject noise during the federated training process, providing privacy guarantees at either the data level (Shokri and Shmatikov~\cite{shokri2015privacy}) or the user level (McMahan et al.\cite{mcmahan2017learning}).
Since the supporting server could also be compromised, some lines of work have focused on concealing individual model updates by performing secure aggregation of such values.
This is achieved using techniques such as secret sharing (SEPIA\cite{burkhart2010sepia}), MHE (Shi et al.\cite{shi2011privacy}, Chan et al.\cite{chan2012privacy}), or additive masking (Bonawitz et al.\cite{bonawitz2017practical}, Bell et al.\cite{bell2020secure}).
Note that secure aggregation is orthogonal to our approach and can potentially be combined with it to conceal the individual updates corresponding to the exposed layers.

To completely prevent leakage from intermediate (aggregated) models, MPC can be employed, allowing the data owners to jointly execute the training mechanism in a secure manner, usually by exploiting secret sharing schemes.
However, a major challenge arises when scaling to a large number of parties, as it leads to impractical communication complexity.
To work around such overhead, the data-owners can delegate the computations to a small cluster of non-colluding servers, usually composed of 2 parties (SecureML~\cite{mohassel2017secureml}), 3 parties (ABY\textsuperscript{3}~\cite{mohassel2018aby3}, Falcon~\cite{wagh2020falcon}, SecureNN~\cite{wagh2019securenn}), or 4 parties (FLASH~\cite{byali2019flash}, Trident~\cite{chaudhari2019trident}).
However, this delegation-based approach imposes strong assumptions on the non-collusion of the computing servers, strongly constraining the threat model.

To overcome the limitations of small cluster MPC solutions to the threat model, a promising research direction has emerged, leveraging FHE schemes to encrypt the model.
By employing FHE, the federated learning process can be conducted entirely under encryption, enabling secure collaboration among a large number of parties
(SPINDLE~\cite{froelicher2021scalable} for generalized linear models, POSEIDON~\cite{sav2020poseidon} for neural networks).
Our solution aligns with this trajectory and can be seen as a generalization of these approaches, with the primary aim of enhancing the efficiency of FL under FHE and making its use feasible in real-world scenarios.

\section{Conclusion and Future Work}
\label{sec:conclusion}

In this paper, we present \tool, a flexible solution for privacy-preserving training of neural networks in a federated setting.
Our system allows users to trade-off little privacy for higher training performance, by selectively encrypting specific portions of the model using a multiparty FHE scheme.
Through an investigation of various privacy attacks in the gray-box setting, where the adversary's access is limited to the unencrypted layers of the model, we determine the layers that tend to leak more information and, consequently, identify which layers are advisable to encrypt.
Our findings indicate that encrypting the last layers is particularly effective to mitigate membership inference attacks, while encrypting the first layers helps preventing model inversion attacks.
In the future, we plan to expand our investigation to include other classes of machine learning models.
Additionally, we aim to enhance our framework to support various optimizers, loss functions, and layer types to make it more applicable to different datasets.
Finally, we believe that our collaborative learning solution could be further optimized performance-wise by adding other MPC techinques to the toolset, for instance to compute the non-linear activation functions of the model.

\section*{Acknowledgment}
This project has received funding from the European Union’s Horizon 2020 research and innovation programme under Grant Agreement No 965315. This result reflects only the author's view and the European Commission is not responsible for any use that may be made of the information it contains.

\bibliographystyle{IEEEtranS}
\bibliography{bibliography}

\appendix

\subsection{MLPs and Gradient Descent}
\label{app:mlp}

Multilayer Perceptrons (MLPs), also known as fully-connected or dense networks, are the simplest kind of feedforward neural networks, where each neuron in one layer is connected to every neuron in the next layer.
Due to their structure, the parameters of these models can be represented by matrices, and in some cases, an additive bias parameter is included for added flexibility.
Given an MLP with $\modelDepth$ layers and $i\in \{1,\dots,\modelDepth\}$, we denote by $\weight_i$ and $\bias_i$ the weight matrix and bias vector between layer $i - 1$ and layer $i$, respectively. 
We denote by $l_i$ the output of layer $i$, that is $l_i = \modelActFunc_i(l_{i-1} \weight_i + \bias_i)$, where $l_0 := x$ and $\modelActFunc_i$ is the so-called activation function used to incorporate non-linearity in the model.
And we denote the intermediate linear application output as $u_i = l_{i-1} \weight_i + \bias_i$.
With a little abuse of notation, we will sometimes refer to the weight and bias $\weight_i, \bias_i$ as to the parameters of the layer they allow to transition to, namely layer $i$.
The model is then a parameterized function $\model(x; \weight_i, \bias_i)$, whose output is $l_\modelDepth$.

The model is trained by minimizing the empirical risk with respect to a given loss function.
For supervised learning, we assume to have a training dataset $\dataset$ of labeled examples $(x,y)$, where $x$ is the feature vector and $y$ is the ground-truth label.
Given a model $\model(x; \weight_i, \bias_i)$, the goal is to optimize the model's parameters by minimizing some loss function $L(\model(x), y)$.
To estimate the gradient of $L$ with respect to the model parameters $\weight_i, \bias_i$, feedforward and backpropagation are used. During feedforward, an input data $x$ is propagated layer by layer through the network, computing all the $u_i$ and $l_i$.
The output prediction $l_\modelDepth$ is then compared to the actual label $y$ using the chosen loss function $L$ to compute the loss value.
The backpropagation algorithm then calculates the gradients of the loss function with respect to the model's parameters.
When the loss function $L$, the model $\model$, and the example $(x, y)$ are clear from the context, we will write $\nabla \weight_i$ and $\nabla \bias_i$ in place of $\nabla_{\weight_i} L(\model(x), y)$ and $\nabla_{\bias_i} L(\model(x), y)$, respectively.

This step is repeated for a batch of examples $B$, and the resulting gradients are averaged to get a better approximation of the actual loss gradient on the real population.
The parameters are then updated by following the negative direction of the gradient, by a step size proportional to a learning rate $\eta > 0$:
\begin{alignat*}{2}
    \weight_i &\leftarrow \weight_i &&- \frac{\eta}{|B|} \sum_{(x, y) \in B}{\nabla_{\weight_i} L(\model(x), y)} \, , \\
    \bias_i &\leftarrow \bias_i &&- \frac{\eta}{|B|} \sum_{(x, y) \in B}{\nabla_{\bias_i} L(\model(x), y)} \enspace .
\end{alignat*}
This iterative process of feeding the data forward, computing the loss, and updating the model's parameters continues until convergence or for a fixed number of iterations.

\subsection{Dataset Description}
\label{app:dataset}
In the following, we give a more detailed description of the datasets we used for our experiments.

\subsubsection*{AT\&T Database of Faces}
This face dataset~\footnote{\url{https://cam-orl.co.uk/facedatabase.html}} was created at the AT\&T Laboratories Cambridge.
It consists of 400 gray-scale images of size 112x92, depicting the faces of 40 individuals in various lighting conditions and facial expressions.

\subsubsection*{EMNIST Letters}
This letter dataset~\footnote{\url{https://www.nist.gov/itl/products-and-services/emnist-dataset}} is part of the extended version of the MNIST dataset by NIST~\cite{cohen2017emnist}.
It consists of 145,600 gray-scale images, representing both upper- and lower-case handwritten letters, which has been centered and resized to 28x28. 
The dataset contains 26 classes, one for each letter from `a' to `z'.

\subsubsection*{Labeled Faces In the Wild (LFW)}
This face dataset~\footnote{\url{http://vis-www.cs.umass.edu/lfw/}} was developed by researchers at the University of Massachusetts, Amherst~\cite{huang2008labeled}.
It consists of 13,233 RGB images, depicting the faces of 5,749 individuals.
The dataset has been further labeled with attributes such as gender, race, age group, hair style, and eyewear.

\subsubsection*{Locations}
This location dataset~\footnote{\label{datasets_link}\url{https://github.com/privacytrustlab/datasets}} was created by the authors of~\cite{shokri2017membership} from Foursquare check-in data for the city of Bangkok.
The processed dataset contains 5010 examples, each corresponding to a unique user.
Each record comprises 446 binary features, indicating whether a user visited a specific region or location type.
The data is clustered into 30 classes, representing different geosocial types.
Following~\cite{shokri2017membership}, we use 1200 examples for training, and the remaining data for validation.

\subsubsection*{MNIST}
Standard handwritten digits dataset by NIST~\footnote{\url{http://yann.lecun.com/exdb/mnist}}.
It consists of 70,000 gray-scale images, centered, and resized to 28x28. The dataset contains 10 classes, one for each digit from '0' to '9'.
Due to the small number of classes and the low feature variability within the same class, this dataset has been observed to be particularly resilient against membership inference attacks~\cite{shokri2017membership}.
For evaluation purposes, we want to start from a situation in which the target model is vulnerable. Thus, we drastically reduce the training set to a mere 100 examples.
For compatibility with our current implementation of \tool, we resize the images to 8x8.

\subsubsection*{Purchase-100}
This purchase dataset~\footnoteref{datasets_link} was created by the authors of~\cite{shokri2017membership} starting from Kaggle's ``acquire valued shoppers'' challenge dataset, containing the shopping history data of several users.
The processed dataset contains 197,324 examples, each corresponding to a unique user.
Each record comprises 600 binary features, indicating whether a user purchased a given product.
The data is clustered into 100 classes, representing different purchase styles.
For our experiments, we use 1000 examples for training and the remaining data for validation.

\subsubsection*{Texas-100}
This hospital dataset~\footnoteref{datasets_link} was created by the authors of~\cite{shokri2017membership} starting from the Hospital Discharge Data records released by the Texas Department of State Health Services.
The processed dataset contains 67,330 examples, each corresponding to a unique patient.
Each record comprises 6,169 binary features, containing information about the patient, the causes of injury, the diagnosis, and the procedures the patient underwent.
The data is clustered into 100 classes, representing the 100 most frequent medical procedures present.
For our experiments, we use 1000 examples for training and the remaining data for validation.

\subsection{Threshold CKKS}
\label{app:threshold_CKKS}

Threshold CKKS supports the same operations as the regular single-key CKKS (see Section 2 of~\cite{kim2022approximate} for details). The operations that are different in the threshold CKKS are public key generation, evaluation key generation (for rotations and/or multiplication), decryption, and bootstrapping. 

The public key generation is done in a distributed manner, where each party generates a public key share (a public key for its secret share) using a common random polynomial, and then all public key shares are summed up to generate the collective public key. The generation of a collective automorphism (rotation) key is done in the same manner because automorphism is a linear operation. The only difference is that multiple common random polynomials may be needed (one for each digit of each automorphism key). For more details, see~\cite{mouchet2021multiparty}.

The generation of the relinearization (multiplication) key is more involved: it requires extra rounds as the encrypted key is a square of the original secret key, and the evaluation of quadratic function has to be done in two steps, as illustrated in Protocol 2 of~\cite{mouchet2021multiparty}.

The decryption is also done in a distributed manner. Each party obtains a partial decryption by evaluating the inner product with respect to their secret share. The partial decryption results are summed up to yield the decryption results. To hide the secret share used for partial decryption from other parties that may see the partial decryption, smudging/flooding noise has to be added during the evaluation of partial decryption~\cite{asharov2012multiparty}. This noise has to be significantly larger than the current approximation noise in CKKS to erase any traces of the secret share. Kluczniak and Santato~\cite{KS23} suggested tight flooding noise estimates for threshold FHE based on the prior estimates for the flooding noise required in the context of approximate homomorphic encryption~\cite{BMSS22}. Note that these estimates are higher than those for achieving IND-CPA$^D$ security for CKKS~\cite{BMSS22}; hence it is sufficient to only consider threshold FHE flooding noise in the case of threshold CKKS. OpenFHE provides a way to configure the flooding noise based on the statistical security parameters specified by the user.

The distributed CKKS bootstrapping procedure performs masked partial decryptions using secret shares and then adds an encryption of the negated mask using large parameters, i.e., $Q_L$, to generate a refreshed encryption of the message. The mask requires additional 2-3 RNS limbs to achieve desired statistical security. The procedure is described in more detail in~\cite{sav2020poseidon}.

\subsection{Homomorphic Operations}
\label{app:he_operations}

To perform homomorphic computations efficiently, we can pack an entire vector in a single ciphertext and exploit the SIMD capabilities of CKKS to perform vector addition and component-wise multiplication in constant time.\footnote{If the vector is too long to fit the number of slots dictated by given FHE parameters, one can split the vector among multiple ciphertexts.}
However, in a neural network, we also need to multiply by weight matrices and evaluate non-linear activation functions.
Here, we describe the approaches we adopt to perform efficient vector-matrix multiplication under encryption and homomorphically evaluate non-polynomial functions.

\subsubsection{Matrix Multiplication}
To efficiently perform matrix multiplication under encryption, the idea is to encode the matrix as a vector in such a way that only one homomorphic multiplication is required.
There exist multiple encoding schemes in the literature for matrices. For instance, in the column-based approach~\cite{kim2018logistic,HHCP18}, one encodes a matrix by concatenating its columns one after the other.
The vector-matrix multiplication is then performed by first replicating the vector to match the number of columns of the matrix, then performing a SIMD multiplication between the two, and finally by performing cumulative addition of the result.
All those operations can be realized by combining homomorphic additions, rotations, and multiplications by a masking vector.
Moreover, the vector replication and the cumulative addition can be made more efficient by recursion, though requiring padding the inputs to a suitable power of two.
Similarly to this column-based approach, a row-based encoding can be employed as well~\cite{BGPRV20}.

To perform subsequent matrix multiplications, we use the alternating packing approach proposed in~\cite{sav2020poseidon}.
It is based on the observation that the result of a vector-matrix multiplication in column-based encoding requires extra rotations to prepare it for a multiplication with another column-based encoded matrix, while it is perfectly ready for a multiplication with another row-based encoded matrix.
The idea is then to encode the matrices that are in consecutive secret layers by alternating between column- and row-based encodings.
Note that multiplying by the transpose of a matrix is equivalent to multiplying by the matrix in the opposite encoding.
We exploit this property to efficiently compute gradients in the back-propagation step.

\subsubsection{Evaluating Non-Polynomial Functions}
We use the Chebyshev interpolation to homomorphically evaluate non-linear functions for a given input range, which is implemented in OpenFHE for the CKKS scheme~\cite{albadawi2022openfhe}. The evaluation of Chebyshev series in OpenFHE is performed using the Paterson-Stockmeyer algorithm adapted to Chebyshev basis~\cite{CCS19}, which requires only $O(\sqrt d)$ homomorphic multiplications to evaluate degree-$d$ polynomials.

\subsection{FHE Micro-benchmark}
\label{app:microbenchmark}

In this section, we provide measurements for various FHE functionalities, enabling estimation of our approach's scalability for different model architectures.
The execution run time and communication size per training party, averaged over multiple runs, are presented in \Cref{tab:pem_microbenchmark}.
For each functionality, we have divided the execution time into computation and communication time.
Note that the missing time from the total execution time reflects the idle time when parties are waiting for other parties to complete their computations in order to proceed.
\begin{table*}[htbp]
    \caption{Microbenchmarks of different FHE functionalities, for $\numParties = 3$ parties, 5-bit integral precision, 55-bit decimal precision, 8-level moduli tower, and $2^{15}$ cyclotomic ring degree. The \textit{one layer} functionality refers to a fully connected layer, while the \textit{one pass} functionality refers to an entire pass (forward or backward) of the model under encryption.}
    \begin{center}
        \renewcommand{\arraystretch}{1.2}
        \begin{tabular}{|l|r|r|r|r|}
        \hline
        \multicolumn{1}{|c|}{Functionality} & \multicolumn{3}{c|}{Execution time (s)} & \multicolumn{1}{c|}{Comm.} \\
        \cline{2-4}
        & \multicolumn{1}{c|}{Comp.} & \multicolumn{1}{c|}{Comm.} & \multicolumn{1}{c|}{Total} & \multicolumn{1}{c|}{size (MB)} \\
        \hline
        Vector-Matrix mult. (64x32) & 6.289 & - & 6.289 & - \\
        Vector-Matrix mult. (32x16) & 3.291 & - & 3.291 & - \\
        Sigmoid evaluation (deg. 13) & 11.046 & - & 11.046 & - \\
        FHE setup & 0.043 & 0.020 & 0.143 & 0.001 \\
        Pub./Priv. key gen. & 0.151 & 0.347 & 0.611 & 12.005 \\
        Relin. key gen. & 0.900 & 0.788 & 2.336 & 81.020 \\
        Rot. key gen. (x23) & 4.378 & 86.458 & 101.662 & 1296.182 \\
        Model generation & 0.989 & - & 0.989 & - \\
        Decryption & 0.376 & 34.090 & 34.999 & 15.013 \\
        Bootstrapping & 1.265 & 49.287 & 51.160 & 22.767 \\
        One layer forward & 20.294 & 147.737 & 170.125 & 68.302 \\
        One layer backward & 24.617 & 98.699 & 124.262 & 45.535 \\
        One pass forward & 60.944 & 443.211 & 510.438 & 204.907 \\
        One pass backward & 73.913 & 296.096 & 372.849 & 136.605 \\
        One training pass & 134.857 & 739.307 & 883.287 & 341.512 \\
        Send model & 0.157 & 17.674 & 17.831 & 27.015 \\
        Update params. & 2.039 & - & 2.039 & - \\
        Aggregation & 1.697 & 56.454 & 60.265 & 37.532 \\
        \hline
        \end{tabular}
    \label{tab:pem_microbenchmark}
    \end{center}
\end{table*}

\subsection{Differential Privacy for Exposed Layers}
\label{app:dp}

To mitigate the leakage from the exposed layers
additional privacy-enhancing techniques can be employed, such as differential privacy~\cite{dwork2006calibrating}.
Applying DP to the parameters or gradients of exposed layers would provide a theoretical privacy guarantee to our solution, albeit introducing an additional trade-off between privacy and accuracy.
By applying noise only on the exposed layers, rather than the entire model, our approach can achieve a higher level of privacy for the same privacy budget compared to standard FL solutions with differential privacy~\cite{shokri2015privacy,mcmahan2017learning,li2019privacy,truex2020ldp}.

We adapt the approach of Shokri and Shmatikov~\cite{shokri2015privacy}, which uses the sparse vector technique~\cite{dwork2014algorithmic,hardt2010multiplicative} to privately upload a small, perturbed subset of the gradients to the global model.
Given a privacy budget $\privacyBudget$ per epoch allocated to each training party $\party_i$, we split this budget among the exposed parameters, and use Laplacian mechanism to add noise to the corresponding gradient value.
The sensitivity of the training mechanism is estimated by clipping the gradient values within the range $[-\clipGradRange,\clipGradRange]$, resulting in a sensitivity of $2\clipGradRange$.
The clipping range value should be independent of the specific training dataset, to avoid leaking sensitive information.
We suggest setting it by calculating the median of the unclipped gradients over the course of training, as proposed in~\cite{abadi2016deep}.

For each value $g$ in $\nabla \weight_j$ for $\layer_j \in \exposedLayers$, random Laplacian noise $\noise_g \sim \laplacianDist(2 \numExposedParams \clipGradRange_g / \epsilon)$ is generated, where $\clipGradRange_g$ is the estimated clipping bound for $g$, and $\numExposedParams$ is the total number of exposed gradients
$\numExposedParams = \sum_{\layer_j \in \exposedLayers}{(|\nabla \weight_j| + |\nabla \bias_j|)}$. The gradient $g$ is then clipped within $[-\clipGradRange_g, \clipGradRange_g]$, and the noise $\noise_g$ is added before uploading to the central server.
A similar process can be followed for the exposed bias gradients.
These operations are performed on the aggregated gradient obtained after several local iterations, each computed over a randomly sampled batch.
Applying noise to each computed gradients could be done as well, and the overall effect over batches can be analyzed using the privacy amplification theorem~\cite{kasiviswanathan2011can,beimel2014bounds}. More advanced techniques, such as privacy accountants~\cite{mcmahan2017learning,abadi2016deep}, can also be potentially adapted to work in our partially encrypted model solution.

As additional remark, differential privacy could also be achieved by reducing the precision of the underlying FHE scheme, thus allowing for larger encryption error on the secret layers.
This error would propagate to the exposed layers during backpropagation, producing a DP-like effect.

\end{document}